%% file: manuscript.tex
\documentclass{jfm}
\usepackage{graphicx}
\usepackage{epstopdf, epsfig}
\usepackage{blindtext}
\usepackage[colorlinks,citecolor = blue, linkcolor=red,hyperindex,CJKbookmarks]{hyperref}
\usepackage{mathtools}
\usepackage{amsmath}
\usepackage{enumitem}
\usepackage{bm}
\input{UDSymbol} 

\makeatletter
 \def\@eqnnum{{\normalfont \color{black} (\theequation)}} \makeatother
 
\shorttitle{Instabilities driven by diffusio-phoretic flow}
\shortauthor{Yibo Chen \textit{et al.}}
\title{Instabilities driven by diffusio-phoretic flow on catalytic surfaces}
\author{
Yibo Chen\aff{1},
Kai Leong Chong\aff{1}\corresp{\email{k.l.chong@utwente.nl}},
Luoqin Liu\aff{1},
Roberto Verzicco\aff{1,2,3} \and
Detlef Lohse\aff{1,4}\corresp{\email{d.lohse@utwente.nl}}
}

\affiliation{
\aff{1}Physics of Fluids Group, Max Planck Center for Complex Fluid Dynamics, MESA+ Institute and J.M.Burgers Center for Fluid Dynamics, University of Twente, P.O. Box 217, 7500 AE Enschede, The Netherlands
\aff{2}Dipartimento di Ingegneria Industriale, University of Rome `Tor Vergata', Via del Politecnico 1, Roma 00133, Italy
\aff{3}Gran Sasso Science Institute - Viale F. Crispi, 7 67100 L'Aquila, Italy
\aff{4}Max Planck Institute for Dynamics and Self-Organisation, 37077 G\"ottingen, Germany}
\begin{document}

\maketitle

\begin{abstract}
We theoretically and numerically investigate the instabilities driven by diffusio-phoretic flow, caused by a solutal concentration gradient along a reacting surface. The important control parameters are the P\'eclet number $Pe$, which quantifies the ratio of the solutal advection rate to the diffusion rate, and the Schmidt number $Sc$, which is the ratio of viscosity and diffusivity. First, we study the diffusio-phoretic flow on a catalytic plane in two dimensions. From a linear stability analysis, we obtain that for $Pe$ larger than $8\pi$, mass transport by convection overtakes that by diffusion, and a symmetry-breaking mode arises, which is consistent with numerical results. For even larger $Pe$, non-linear terms become important. For $Pe > 16\pi$, multiple concentration plumes are emitted from the catalytic plane, which eventually merge into a single larger one. When $Pe$ is even larger ($Pe \gtrsim 603$ for Schmidt number $Sc=1$), there are continuous emissions and merging events of the concentration plumes. The newly-found flow states have different flow structures for different $Sc$: For $Sc\ge1$, we observe the chaotic emission of plumes, but the fluctuations of concentration are only present in the region near the catalytic plane. In contrast, for $Sc<1$, chaotic flow motion occurs also in the bulk. In the second part of the paper, we conduct three-dimensional simulations for spherical catalytic particles, and beyond a critical P\'eclet number again find continuous plume emission and plume merging, now leading to a chaotic motion of the phoretic particle. Our results thus help to understand the experimentally observed chaotic motion of catalytic particles in the high $Pe$ regime.
\end{abstract}


\section{Introduction}\label{sec:intro}
Self-propulsion at the micrometer scale frequently occurs in nature \citep{lauga2009hydrodynamics, lauga2016bacterial, bray2000cell, jeanneret2016entrainment}. For example, micro-organisms self-propel to search for nutrients, different temperatures, or sunlight. Inspired by such motile biological organisms, extensive studies on artificial micro-swimmers have been done over the last one and a half decades, especially on self-propelled phoretic particles \citep{jiang2010janus, moran2017phoretic, gol2007, kai2020squimer, maass2016swimming, bar2019self, jin2017chemotaxis}. Also dissolving or chemically reacting droplets can show such phenomena \citep{kruger2016curling, li2019bouncing, babak2019, lohse2020}. A typical feature of the self-propelled particles is that, instead of swimming with appendages, they can propel themselves by converting free energy from the environment into kinetic energy \citep{ramaswamy2010mechanics, ebbens2010pursuit}.

\begin{figure}
\centering
\centering \includegraphics[width=0.8\textwidth]{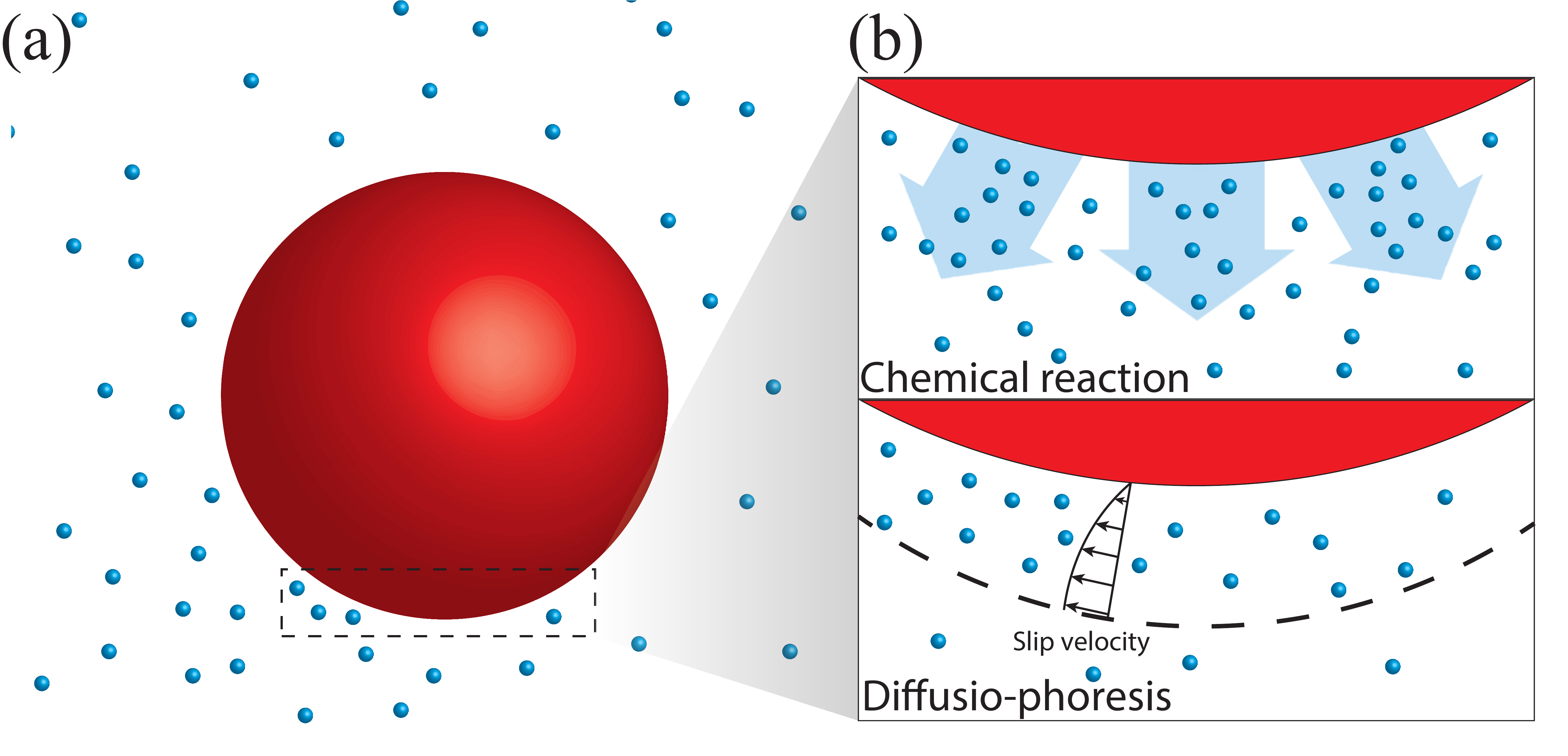}
\caption{Schematic illustration of the catalytic particles (red) with chemical reaction and diffusio-phoresis near the interface. The product originating from the catalytic reaction at the particle surface is shown in cyan. (b) shows a zoom-in of (a). If there is a concentration gradient at the interface, a slip velocity is induced (diffusio-phoresis). Beyond a critical reaction rate (expressed as a critical P\'eclet number), such gradient emerges through a linear instability.}
\label{fig:fig1}
\end{figure}

The driving mechanism behind the propulsion of phoretic particles is diffusio-phoresis \citep{anderson1989colloid}. Note that in some literature the terminology ``diffusio-osmotic effect" is used to indicate the same mechanism. The basic feature is that whenever there exists a tangential concentration gradient on the surface of the particle, there is an induced flow within the interaction layer adjacent to the surface, as shown in figure \ref{fig:fig1}. Since the layer is much thinner than the size of the object, the flow is conveniently described with a slip velocity at the surface \citep{gol2007}. This effect can also be generalized to other coupled fields such as the temperature or the electric fields. The resulting flows are respectively referred to as thermo-phoretic or electro-phoretic \citep{piazza2008thermophoresis, squires2006breaking, long1999electroosmotic, moran2011electrokinetic}. 

The classical mathematical framework for the study of self-propelled particles has often neglected the effect of solute advection \citep{gol2007}. \cite{michelin2013spontaneous}, however, have revealed that the P\'eclet number $Pe$ is an important parameter controlling the motion of self-propelled particles. $Pe$ is the ratio of the solute advection to the diffusion rates. Through a linear stability analysis and corresponding simulations, \cite{michelin2013spontaneous} found that when $Pe$ is larger than the critical value $Pe_{cr}=4$, a spherical active particle by dissolution and chemical reaction exhibits a motion in a preferred direction which breaks the rotational symmetry of the system. Later, \cite{michelin2014phoretic} performed a comprehensive theoretical study on how the moving speed of the active particle depends on $Pe$, and generalized the theory to any coverage of the reacting surface.

For large enough $Pe$, some fascinating features can emerge. \cite{PhysRevLettHu} have numerically observed that for large enough P\'eclet numbers, such an active isotropic particle acquires chaotic trajectories. Analogously, in the problem of active droplets, \cite{ruckenstein1981can} also found similar helical or chaotic motions, as caused by the interfacial Marangoni flow \citep{suga2018self, maass2016swimming, herminghaus2014interfacial}. Though the phenomena of active droplets and active particles look similar, \cite{kruger2016curling} explained that the helical trajectory of the active droplet is attributed to the coupling between the internal flow and the direction of the nematic field, whereas such internal flow is obviously absent in particles. In a further study, \cite{morozov2019nonlinear} have considered both Marangoni and diffusio-phoretic effects into their numerical simulation, and also demonstrate that a chaotic oscillation of the droplet can occur.

\begin{figure}
\centering \includegraphics[width=0.5\textwidth]{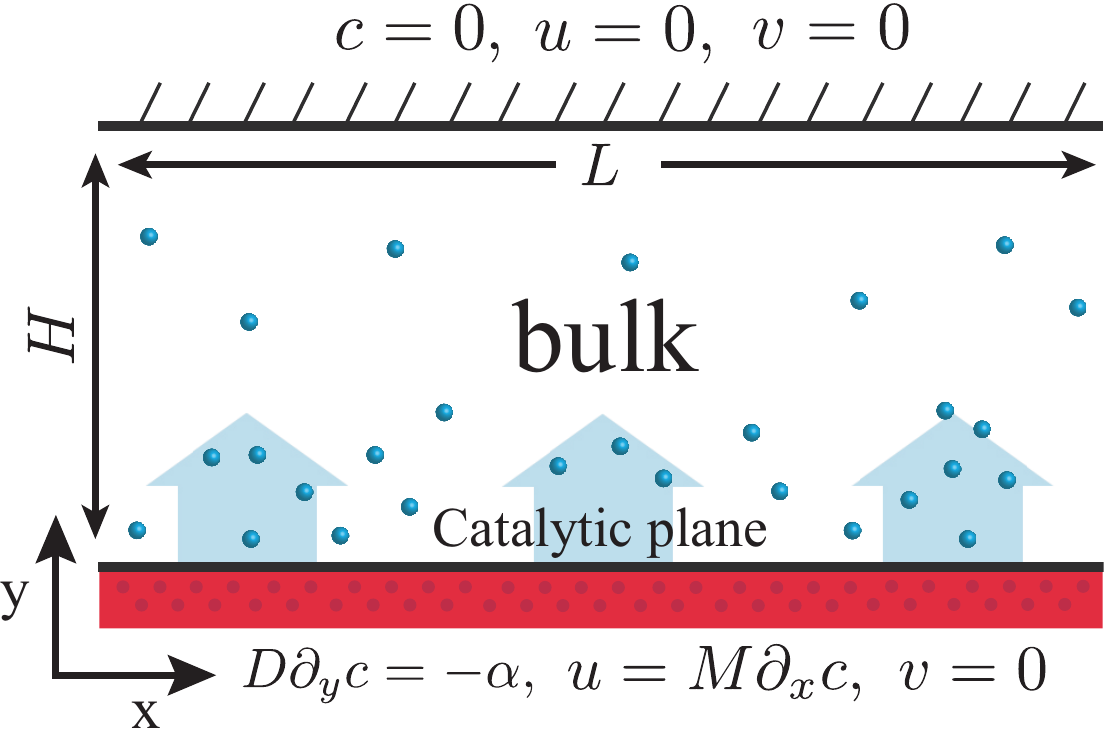}
\caption{The setup of the system with the boundary conditions. A catalytic plane is located at the bottom of the domain. Periodic boundary conditions are applied in $x$-direction.}
\label{fig:fig2}
\end{figure}

Very recently, \cite{michelin2020spontaneous} investigated a simplified system, namely a uniform phoretic channel. They reported spontaneous symmetry-breaking of the solute distribution which provides a route to understand the propulsion of isotropic active particle. However, it remains necessary to understand how the Schmidt number (defined as the ratio of kinematic viscosity to solute diffusivity) influences the diffusio-phoretic instability. Furthermore, the high P\'eclet number regime is still not fully explored, and we will see that there an interesting chaotic flow arises.

A flow very related to the diffusio-phoretic flow is B\'ernard-Marangoni convection, where a spontaneous flow instability occurs too. In that system, the flow is driven by a surface tension difference caused by a variation of the temperature at the fluid surface \citep{pearson1958convection, davis1987thermocapillary, bergeon1998marangoni, boeck2002low}. These two systems share some similarities in their symmetry-breaking mechanism and with the chaotic flow motion at high enough P\'eclet or Marangoni number. However, the two systems are different problems with diffusio-phoretic flow being driven by the phoretic velocity at the surface and B\'enard-Marangoni being driven by the difference in surface tension.

Motivated by the above mentioned recent findings, in this paper we focus on the instability due to chemical reactions and the resulting diffusio-phoretic flow near a catalytic interface, especially in the large $Pe$ regime. To start with some reduced complexity, we first consider a  simplified model, namely diffusio-phoretic flow over a catalytic plane, in order to study the dynamics near the catalytic surface (see figure $\ref{fig:fig2}$). This simplified model can reproduce the important features of the diffusio-phoretic flow, and it is also convenient to avoid the added complexities arising from the curvature of the surface. In the second part of the paper we go beyond the simplified model and numerically examine the plume emission and merging phenomena for chaotically-moving phoretic particles.

The paper is organized as follows: After a description of the problem setup and the control parameters in Section \ref{sec:modelcat}, the linear stability analysis for the catalytic plane system is performed in Section \ref{theory}. Then the numerical method and numerical setup are provided in Section \ref{nummethod}. The numerical results for the catalytic plane are presented in Section \ref{linearsim}. Then we extend our research to phoretic particle in Section \ref{particle}. Finally, conclusions and outlook to further work are given in Section \ref{sec:conc}. The details of the linear stability analysis is given in Appendix \ref{Aappendix}.

\section{Problem setup and control parameters} \label{sec:modelcat}
We start with the two-dimensional system sketched in figure \ref{fig:fig2}. The domain has periodic boundary conditions on both sides and a catalytic plane at the bottom. The width and height of the domain are denoted by $L$ and $H$. The physical variables to describe the system are the concentration of the product $\hat{c} (x,y,t)$ and the velocity of the fluid $\hat{\boldsymbol{u}}(x,y,t)$. Note that all dimensional physical variables are marked with hat (e.g. $\hat{c}$, $\hat{\boldsymbol{u}}$), while the dimensionless ones without (e.g. $c$, $\boldsymbol{u}$). At the catalytic surface, chemical reactions take place which convert the reactant into the product. By assuming a constant reaction rate, the concentration boundary condition of the product at the bottom plane is given by

\begin{equation} \label{alpha}
\left. D\frac{\partial \hat{c}}{\partial \hat{y}}\right|_{y=0}=-\alpha,
\end{equation}
where $D$ is the diffusivity of the product in the fluid and $\alpha$ measures the strength of the reaction activity at the catalytic surface, i.e. the generation of solute by the reaction. 

The tangential concentration gradient induces a slip velocity at the surface of the plane. This is the so-called diffusio-phoretic flow, which is parallel to the surface and its magnitude is proportional to the tangential concentration gradient. The relationship between the induced slip velocity and the tangential concentration gradient is given by
\begin{equation} \label{eqj1}
\hat{u}|_{y=0}=M\frac{\partial \hat{c}}{\partial \hat{x}},
\end{equation}
where $M$ is the phoretic mobility. The sign of $M$ can either be positive or negative, depending on the type of the solute-surface interaction \citep{anderson1989colloid}.  \cite{michelin2013spontaneous} prove that the diffusio-phoretic system is unstable only if $M\alpha$ is positive. In this work, we study the case $M>0$ and $\alpha>0$.

The time evolution of the concentration field $c(x,y,t)$ and the velocity field $\boldsymbol{u}(x,y,t) =(u(x,y,t),v(x,y,t))$ are governed by the Navier-Stokes equations and the convection-diffusion equation. The characteristic scales for non-dimensionlization are $M \alpha/D$ for velocities, $L$ for lengths and $\alpha L/D$ for concentrations. The dimensionless form of the governing equations can then be written as:
\begin{equation}\label{eqj2}
\frac{\partial c}{\partial t}+\boldsymbol{u}\cdot\nabla c= \frac{1}{Pe}\nabla^2 c,
\end{equation}
\begin{subeqnarray}\label{eqj4}
\gdef\thesubequation{\theequation \mbox{\textit{a}},\textit{b}}
\frac{\partial \boldsymbol{u}}{\partial t}+(\boldsymbol{u}\cdot\nabla)\boldsymbol{u}=-\nabla p+\frac{Sc}{Pe}\nabla^2 \boldsymbol{u},\quad
\boldsymbol{\nabla} \cdot \boldsymbol{u}=0,
\end{subeqnarray}
\returnthesubequation
where $Pe$ is the P\'eclet number,  characterizing the ratio of the solutal advection rate to the diffusion rate and $Sc$ the Schmidt number, characterizing the ratio between the momentum and mass diffusivities:
\begin{subeqnarray}\label{eqj4_1}
\gdef\thesubequation{\theequation \mbox{\textit{a}},\textit{b}}
Pe=\frac{M\alpha L}{D^2}, \quad
Sc=\frac{\nu}{D}.
\end{subeqnarray}
\returnthesubequation
 Note that for the case of fixed $Pe$ and $Sc \rightarrow \infty$, the pressure term in equation (\ref{eqj4}a) should be rescaled as $p'=\frac{p}{Sc/Pe}$, since the pressure gradient always exists, even in Stokes flow, where it then balances the viscous forces.

In dimensionless form, the concentration boundary condition at the catalytic plane becomes
\begin{equation}\label{eqj5_1}
\frac{\partial {c}}{\partial {y}}{\bigg{|}}_{y=0}=-1.
\end{equation}
The tangential velocity is proportional to the tangential concentration gradient, and its dimensionless form is

\begin{equation}\label{eqj5}
u|_{y=0}=\frac{\partial c}{\partial x},
\end{equation}
while the normal component of the velocity vanishes at the plane surface
\begin{equation}\label{eqj6}
v|_{y=0}=0.
\end{equation}
Both the velocity and the concentration boundary conditions at the top wall are zero,
\begin{subeqnarray}\label{eqj8}
\gdef\thesubequation{\theequation \mbox{\textit{a}},\textit{b}}
\boldsymbol{u}|_{y=H}=0, \quad
c|_{y=H}=0.
\end{subeqnarray}
\returnthesubequation

In Section \ref{theory}, we have conducted the linear stability analysis with a semi-infinite domain. We note that for aspect ratios $H/L > 0.8$ (discussed in Section \ref{sim}), the growth rate of the instability becomes insensitive to the aspect ratio. Therefore, we can compare the results on a linear stability analysis for a semi-infinite domain to the numerical results for a finite domain with $H/L=1$.

\section{Linear stability analysis for catalytic plane} \label{theory}
In this section, the linear stability analysis is performed to investigate the stability of the system. In the linear stability analysis we add small amplitude perturbations to the basic state:
\begin{equation} \label{eqj13}
\boldsymbol{u} = \bar{\boldsymbol{u}}(y,t)+\epsilon \tilde{\boldsymbol{u}}(x,y,t), \quad
p =\bar{p}(y,t)+ \epsilon \tilde{p}(x,y,t), \quad
c = \bar{c}(y,t) + \epsilon \tilde{c}(x,y,t),  
\end{equation}
where $\bar{\boldsymbol{u}}$,  $\bar{p}$ and $\bar{c}$ are the basic state of the velocity, pressure and concentration fields, and $\epsilon \tilde{u}, \epsilon \tilde{p}$ and $\epsilon \tilde{c}$ are small perturbations with the coefficient $\epsilon \ll 1$.

A trivial solution to the basic configuration is a static state with zero velocity and pressure. Substitute zero velocity into (\ref{eqj2}), the concentration field is (see Appendix \ref{Aappendix}, see also \cite{Wu2006}, p. 144)
\begin{equation} \label{eqj12}
\bar{c}(y,t)=\int_0^t \frac{ 1 }{\sqrt{\pi Pe (t-\tau)}} \exp \left[ - \frac{Pe y^2}{4 (t-\tau)} \right] \textrm{d} \tau.
\end{equation}
For $t\rightarrow \infty$ and any finite $y$, we obtain the following concentration gradient
\begin{equation} \label{eq.c0-3}
\frac{\partial \bar{c}}{\partial y} = -1
\end{equation}
at the catalytic plane.

Substituting equations (\ref{eqj13}) and (\ref{eq.c0-3}) into the governing equations (\ref{eqj2}) and (\ref{eqj4}), with base flow $\bar{\pu}(y,t)=0$ and $\bar{p}(y,t)=0$, and remain only the $O(\epsilon)$-terms, we get the linearized governing equations are:
\begin{equation}\label{eqj9}
\frac{\partial \widetilde{c}}{\partial t}=-\widetilde{v}+ \frac{1}{Pe}\nabla^2 \widetilde{c},
\end{equation}
\begin{subeqnarray}\label{eqj11}
\gdef\thesubequation{\theequation \mbox{\textit{a}},\textit{b}}
\frac{\partial \boldsymbol{\widetilde{u}}}{\partial t}=-\nabla \widetilde{p}+\frac{Sc}{Pe}\nabla^2 \boldsymbol{\widetilde{u}}, \quad
\boldsymbol{\nabla} \cdot \boldsymbol{\widetilde{u}}=0.
\end{subeqnarray}
The boundary conditions become:
 \beq\lb{bc12}
	\left. \fr{\pat \widetilde{c}}{\pat y} \right|_{y=0} = 0, \quad 
	\left. \widetilde{u} \right|_{y=0} = \left. \fr{\pat \widetilde{c}}{\pat x} \right|_{y=0}, \quad 
	\left. \widetilde{v} \right|_{y=0} = 0. 
     \eeq 
We now assume as ansatz a separation of variables and periodic behavior in the lateral direction, such that the perturbation can be written as:
\begin{equation} \label{eqj14}
(\widetilde{u}(x,y,t), \widetilde{v}(x,y,t),  \widetilde{p}(x,y,t),\widetilde{c}(x,y,t))=(\check u(y), \check v(y), \check p(y), \check c(y))e^{ikx + st},
\end{equation}
where $k=2\pi n$ and $n \in \mathbb{N}$ is the wavenumber. This is the standard normal mode analysis (see, for example, \cite{drazin2004}). Note that the sign of $s$ determines the flow stability of the system: $s>0$ means exponential growth, or instability (the larger, the more unstable), whereas $s<0$ indicates stability.
\begin{figure}
\centering \includegraphics[width=1.0\textwidth]{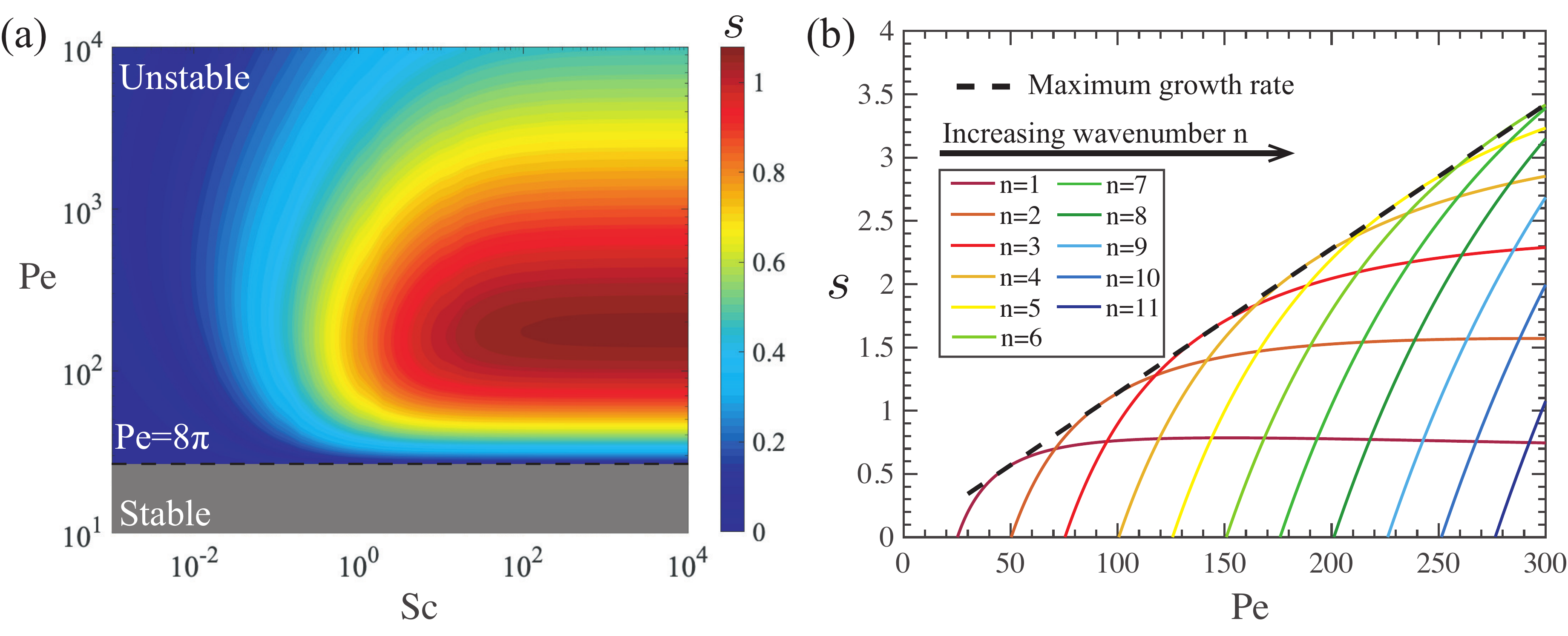}
\caption{\label{fig:fig3}(a) Stability diagram for the catalytic plane in the $Pe$ vs $Sc$ parameter space for wavenumber $n=1$. An eigenvalue $s >0$ indicates instability. The color represents the actual value of $s$, i.e., the strength of the exponential growth. When $Pe>8\pi$, $s$ is positive and the system is unstable, independently of $Sc$. (b) $s$ as a function of $Pe$ at various wavenumber for $Sc=1$ by linear stability analysis. The wavenumber of the curve increases from left to right. For wavenumber $n$, when $Pe<8\pi n$, $s$ is negative and the system is stable. When $Pe>8\pi n$, $s$ becomes positive and the system becomes unstable towards this mode $n$. The function of the maximum growth rate curve (dashed line in (b)) is equation (\ref{eqjenvelop}).}
\end{figure}
Combining the above equations (\ref{eqj9})-(\ref{eqj14}) and the boundary conditions, we obtain the relation which allows us to calculate how the stability depends on $Pe$ and $Sc$ (detailed derivations are in Appendix \ref{Aappendix}):
\begin{equation}
\label{eqj14_5}
Pe  = k \sqrt{1+ \frac{ s Pe }{ k^2}} \left( 1 + \sqrt{1+ \frac{ s Pe }{ k^2}} \right)  \left( \sqrt{1+ \frac{ s Pe }{ Sc\, k^2}} + \sqrt{1+ \frac{ s Pe }{ k^2}} \right). 
\end{equation}

Assuming $s=0$ in equation (\ref{eqj14_5}), we get the critical $Pe$  for transition from stability to instability for different wavenumber $n$,
\begin{equation}
\label{eqjonset}
 Pe_{cr}=4k=8\pi n,
 \end{equation}
Note that $Pe_{cr}$ is independent of $Sc$.

If we combine equation (\ref{eqj14_5}) and its derivative with respect to $n$, we obtain the function of the maximum growth rate at different wavenumber for $Sc=1$ (for a detailed derivation see the Appendix \ref{Aappendix}):
\begin{equation}
\label{eqjenvelop}
s=\frac{85 \sqrt{17} - 349}{128}Pe\approx 0.0114Pe,
 \end{equation}
and the corresponding wavenumber is
\begin{equation}
\label{eqjdomn}
n_{\text{max}}=\lfloor \frac{ 31 - 7 \sqrt{17} }{32\pi} Pe \rceil \approx \lfloor 0.0213Pe \rceil,
 \end{equation}
where the symbol $\lfloor  \ \rceil$ represents the calculation of the nearest integer.
 
The exponential growth rate $s$ as obtained from equation (\ref{eqj14_5}) as function of $Pe$ and $Sc$ for the case of $n=1$ is shown in figure \ref{fig:fig3}(a). Moreover, $s$ as a function of $Pe$ for different wavenumbers and $Sc=1$ is plotted by the solid curves in figure \ref{fig:fig3}(b). The dashed line shows the maximum growth rate curve, which is equation (\ref{eqjenvelop}). The way to calculate figure \ref{fig:fig3} from equation \er{eqj14_5} is explained in Appendix \ref{Aappendix}. 

The linearized diffusion-convection equation (\ref{eqj9}) helps us to understand the physical mechanism of the diffusio-phoretic instability. If there is local concentration variation at the surface, the diffusion term $\frac{1}{Pe}\nabla^2c$ smoothes out the local concentration difference, which makes the system stable. In contrast, dominance of the advection term $-\boldsymbol{u}\cdot\nabla c$ will increase the concentration difference, such that the system becomes unstable. Thus it can be seen that the competing mass transport by diffusion and advection determines the instability, which is quantified by $Pe$. If $Pe$ is above a critical value, the advection term results in positive feedback, which amplifies the disturbance and leads to the instability. 

\begin{figure}
\centering \includegraphics[width=0.6\textwidth]{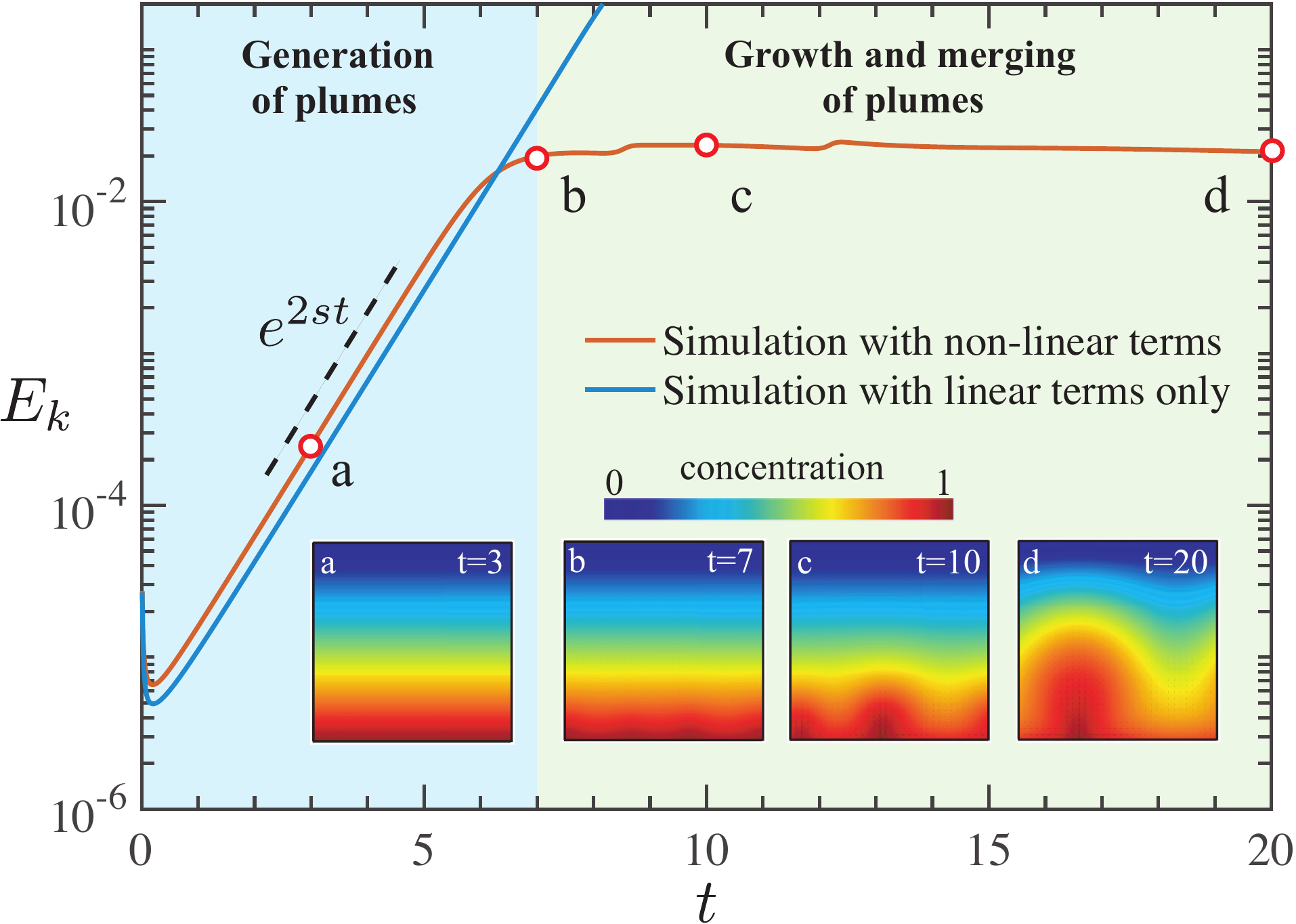}
\caption{\label{fig:fig4} Time evolution of the kinetic energy $E_k$ for  the case $Pe=125$ and $Sc=1$ with random perturbation from simulations with only linear terms (blue solid curve) and with both linear and non-linear terms (red solid curve). The kinetic energy $E_k$ is in log scale. For the case with both linear and non-linear terms, the growth of $E_k$  levels off near time instant b, compared to that with only linear terms, because of non-linear saturation. The process is divided into two subprocesses: plume generation and plume growth and merging. During the first subprocess, $E_k$ grows exponentially $E_k \sim e^{2st}$. The points at the curve represents four states in the process, of which the concentration fields are shown in panels a to d, respectively. (a) Plume generating: Triggered by a perturbation, the kinetic energy increases exponentially. (b-d) Plume growing and merging. As $E_k$ reaches around 0.02, the kinetic energy reaches a plateau; at the same time the plumes emerge in the concentration field. The plumes grow and merge with each other. In the end, only one major plume remains in the field (d).}
\end{figure}

\begin{figure}
\centering \includegraphics[width=0.9\textwidth]{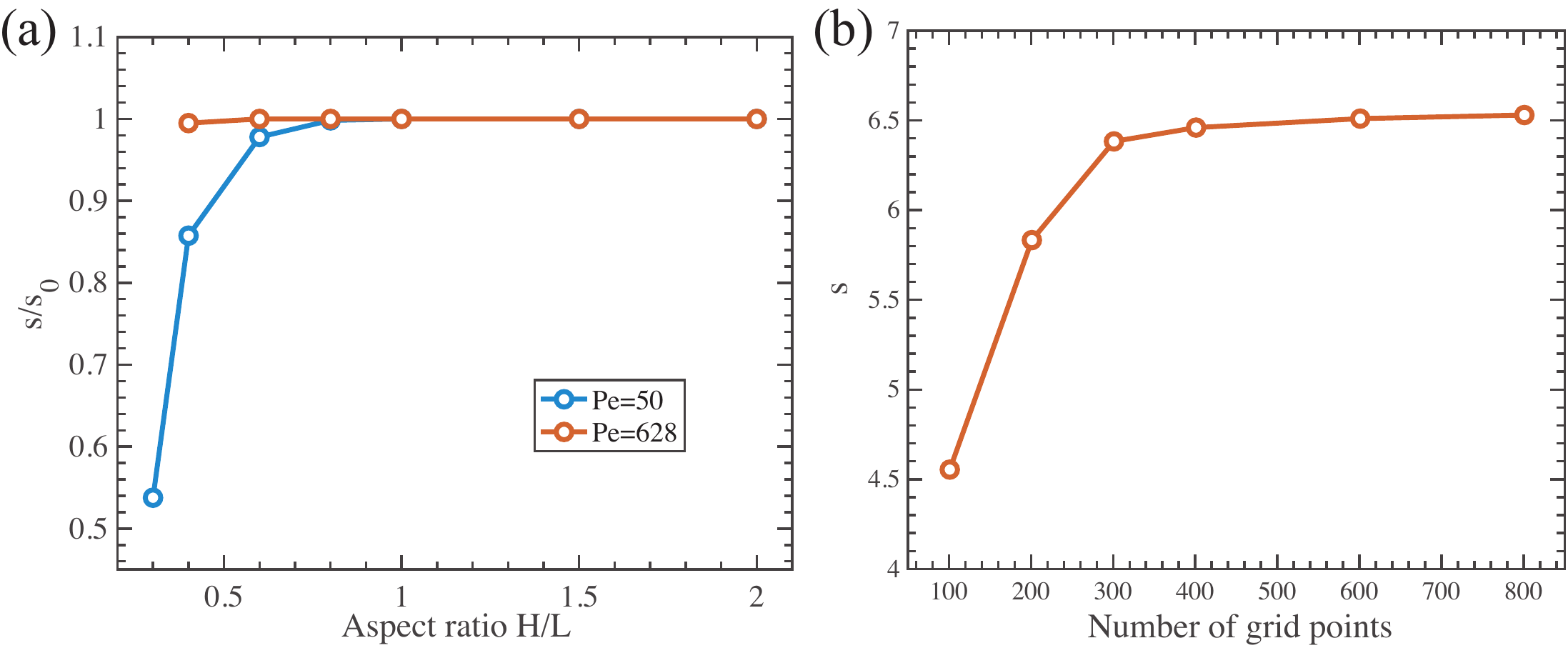}
\caption{\label{fig:fig4height} (a) Normalized growth rate of the dominant unstable mode $s/s_0$ for $Pe=50$ and $628$ with different aspect ratio $H/L$, where $s_0$ is the growth rate obtained at $H/L=2$. It can be seen that when $H/L\geq0.8$, the growth rate becomes insensitive to the aspect ratio. (b) Mesh refinement test with growth rate $s$ versus the number of grid points in one dimension. For the case of Pe=628, the percentage change of $s$ is less than $1 \%$ when the grid resolution increases from $401 \times 401$ to $801 \times 801$.}
\end{figure}

\section{Simulation of catalytic plane} \label{sim}
We now numerically study the diffusio-omostic instability. The objective of the numerical simulaiton is to understand the effect of the non-linear terms and random initial perturbation which are ignored in the linear stability analysis. 

\subsection{Numerical setups}\label{nummethod} 
The fluid motion and concentration field  are solved using direct numerical simulation (DNS) of the Navier-Stokes equations and diffusion-convection equation in Cartesian coordinates.  Equations (\ref{eqj2})-(\ref{eqj4}) are spatially discretized using the central second-order finite difference scheme. Along both horizontal and vertical directions, homogenous staggered grids are used. The equations are integrated by a fractional-step method with the nonlinear terms computed explicitly by a low-storage third-order Runge-Kutta scheme and the viscous terms computed implicitly by a Crank-Nicolson scheme \citep{verzicco1996finite,van2015pencil}. The simulations are then conducted with the concentration and the velocity boundary conditions written in equations (\ref{eqj5_1})-(\ref{eqj8}). We first examine how the growth rate responds to the domain size. Figure \ref{fig:fig4height}(a) shows that when the aspect ratio $H/L\geq0.8$, the growth rate of the instability becomes insensitive to the aspect ratio. Besides, the mesh refinement test is given in figure \ref{fig:fig4height}(b), from which we see the convergence of the growth rate when the number of grid points in one direction has reached roughly $300$. Therefore, we chose $H/L=1$ and the mesh $401 \times 401$ for all our phoretic channel simulations.

The initial condition is the fluid at rest and a constant concentration gradient along the $y$-direction (see equation (\ref{eq.c0-3})). Then a small sinusoidal perturbation is added to the concentration field to trigger the instability:
\begin{equation}
\label{eqj15_0}
\delta c= 10^{-4} \sin (2\pi nx),
\end{equation}
where $n$ is the wavenumber of the perturbation.

\begin{figure}
\centering \includegraphics[width=0.6\textwidth]{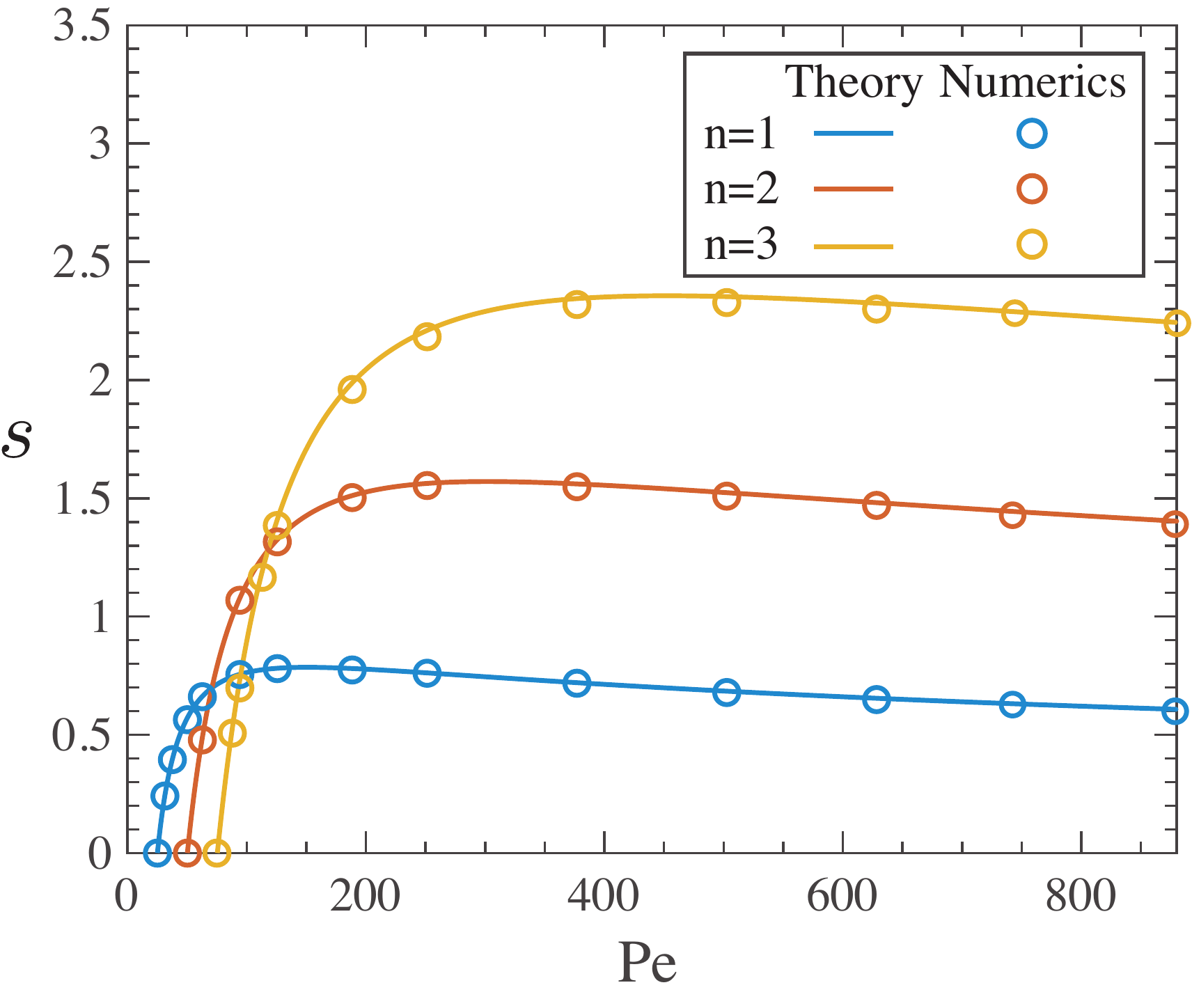}
\caption{\label{fig:fig5} Theoretical (solid lines) and numerical results (circles) of growth rate $s$ for different wavenumber $n=1,2,3$ and $Sc=1$. The simulations are performed with only linear terms.}
\end{figure}

\subsection{Nonlinear saturation}\label{linearsim} 
To quantify the long term growth of the instability, we examine how the kinetic energy $E_k=\frac{1}{A}\int_A \frac{v^2}{2} dS$ ($A$ is the whole domain and $v$ is the velocity) changes in time. An example time series of $E_k$ is shown in figure \ref{fig:fig4}, which corresponds to the case of $Sc=1$ and $Pe=125$. The result suggests that after the initial perturbation, there is a transient stage during which the kinetic energy grows exponentially, i.e. $E_k \sim e^{2st}$. As a consistency test, our simulation confirms that the involvement of non-linear terms in the simulation does not change the initial growth rate $s$. However, later the growth of $E_k$ begins to level off after some time because of non-linear saturation. Such non-linear saturation is common in most linearly unstable non-linear systems, such as Rayleigh-B\'enard convection \citep{greenside1984nonlinear}, Taylor-Couette flow for inner cylinder rotation \citep{Gro2016}, or Rayleigh-Taylor instability \citep{haan1989onset}. The concentration fields at different time show that during the saturated stage, the emitted plumes merge into a larger one, and eventually the flow structure develops into the state with a single large plume. 

Next, we compare the exponential growth rate $s$ of the instability for various wavenumber cases ($n=1,2,3$) during the initial stage with exponential growth as shown in figure \ref{fig:fig5}. For the benchmark cases with only the linear terms, the data points (circles) agree excellently with the linear stability analysis (solid curves). This result can be regarded as further validation for our numerical code. 

\begin{figure}
\centering \includegraphics[width=0.95\textwidth]{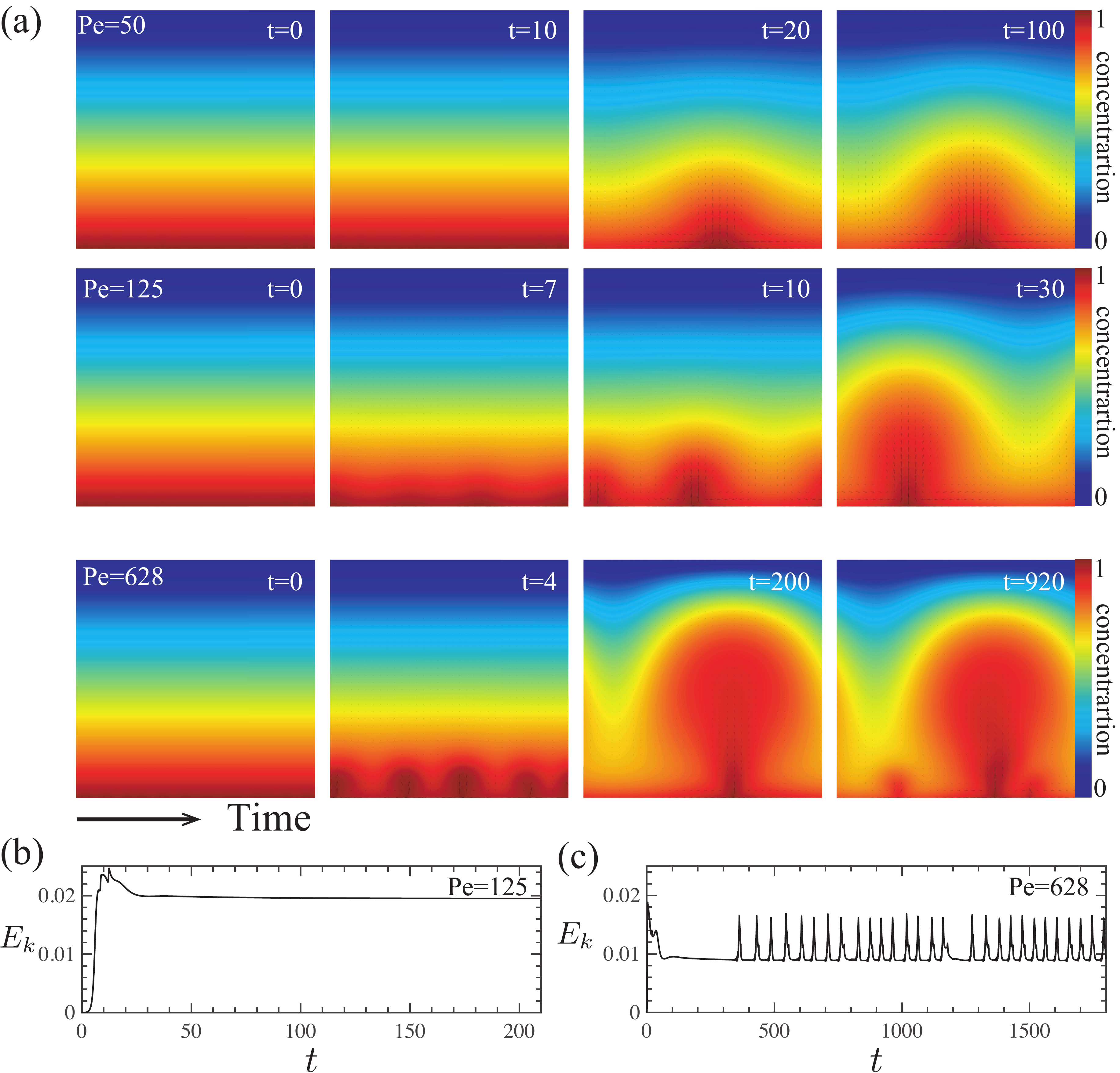}
\caption{\label{fig:fig7} (a) The concentration contours for different $Pe$ numbers: $Pe=50$, $Pe=125$, and $Pe=628$. The simulations are based on random initial perturbation and performed with the full equations, including the non-linear terms. Four snapshots in time are plotted for each $Pe$. First column: beginning state; second and third column: intermediate states; final column: final (statistical) stable state. (b-c) Time evolution of the total kinetic energy of the velocity field for $Pe= 125$ (b) and $628$ (c). For the case of $Pe=125$, the kinetic energy in the final stage converges, while for $Pe=628$ it shows spiky and intermittent signals.}
\end{figure}

We further examine the situation with random initial perturbation solved with the full equations, including nonlinear terms. The above theoretical analysis has shown that for higher $Pe$, the larger wavenumber mode can be triggered. The concentration fields in figure \ref{fig:fig7}(a) provide more insight into the triggering of higher-order modes for larger $Pe$. Different time instants of the concentration fields for different $Pe$ are shown in the figure. For $Pe=50$, there is a single concentration plume generating initially. However, for larger $Pe=125$, multiple plumes are initially emitted. They undergo a merging process to form a single large plume. After formation of the single large plume, $E_k$ reaches the asymptotic value shown in the time series in figure \ref{fig:fig7}(b). Interestingly, for even larger $Pe=628$, $E_k$ has spiky signals within a statistically steady state as shown in figure \ref{fig:fig7}(c). The corresponding concentration fields in figure \ref{fig:fig7}(a) reveals that small plumes are continuously generated from the reacting wall, and the merging of the plumes occurs simultaneously. Such continuous plume emission and merging can also clearly be seen from the Supplementary Movie.

\begin{figure}
\centering \includegraphics[width=0.7\textwidth]{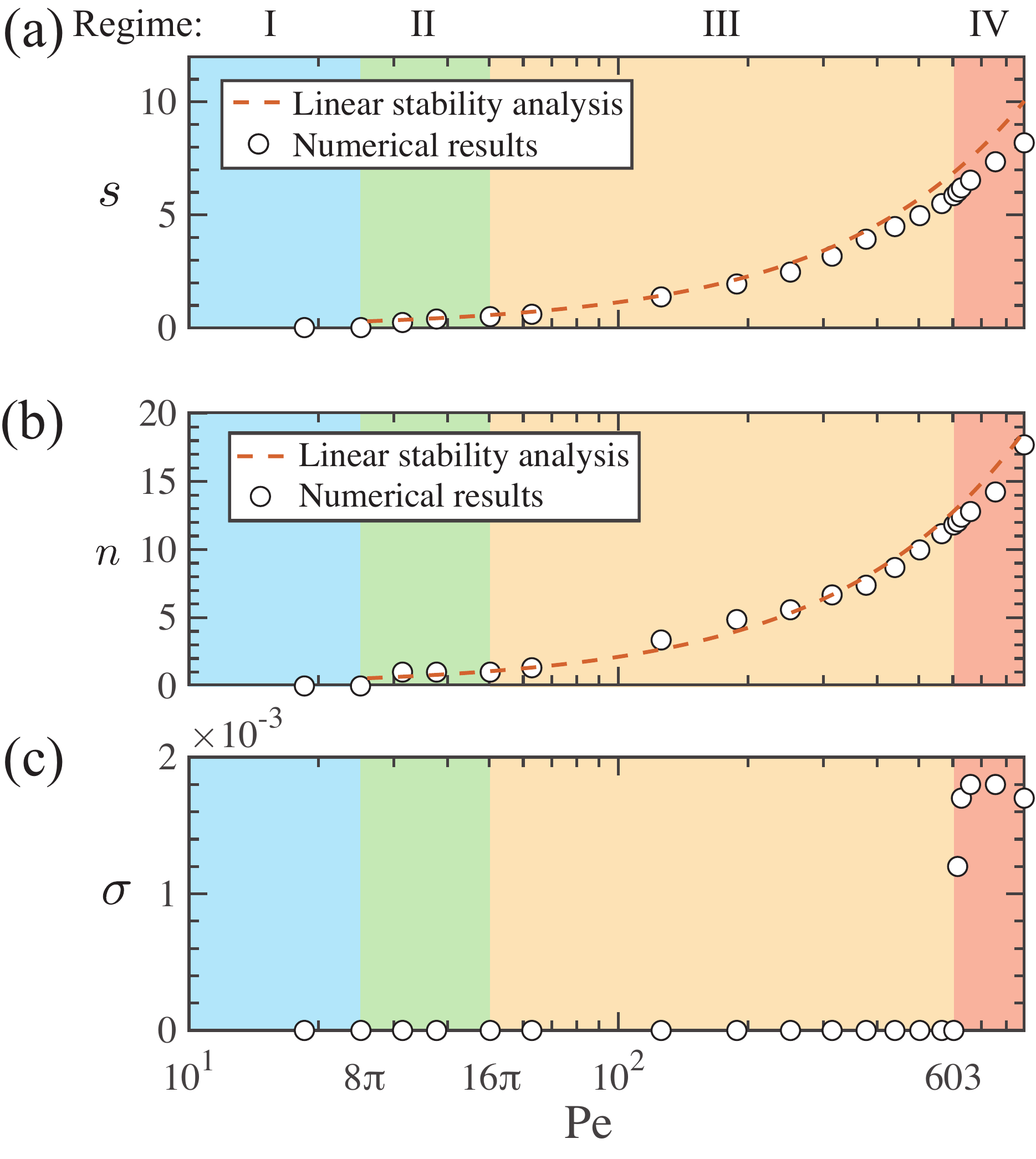}
\caption{\label{fig:fig8} The simulation result for the catalytic plane with non-linear terms and random initial perturbation for $Sc=1$ and different $Pe$: (a) Theoretical (dashed curve, which is equation (\ref{eqjenvelop})) and numerical result (circle) of $s$ as a function of $Pe$, which indicates that the system becomes unstable when $Pe>8\pi$. (b) Theoretical (dashed line, equation (\ref{eqjdomn}) without rounding operation) dominant wavenumber and numerical wavenumber $n$ calculated by Fourier transform (circle) as a function of $Pe$. The result indicates that when $Pe>16\pi$, multiple plumes are generated. (c) Standard deviation $\sigma$ of the kinetic energy for different $Pe$, which indicates that when $Pe \gtrsim 603$, the kinetic energy eventually fluctuates because small plumes are continuously generated. Thus four regimes are classified, marked with different colors: stable (I: blue), a single wave (II: green), multiple waves which merge with each other (III: orange), and multiple waves with small plumes continuously being regenerated (IV: pink).}
\end{figure}

Finally, we classify the four regimes based on the three criteria: 
	\begin{itemize}[leftmargin=*,labelindent=6mm,labelsep=3mm]
		\item Growth rate of the instability.
		\item Number  of plumes generated initially.
		\item Fluctuation of the kinetic energy ($E_k$) after reaching the statistically steady state.
	\end{itemize}

To quantify the number of generated plumes in the initial stage, we perform a Fourier transformation of the concentration field along the reacting wall ($y=0$) at the instant when the plumes emerge (instant b in figure \ref{fig:fig4}). The wavenumber, i.e., the initial number of plumes (circles), is compared with the dominant wavenumber as obtained from linear stability analysis (red dashed line) in figure \ref{fig:fig8}(b). Both are in good agreement. The dominant wavenumber is that of the maximum growth rate $s$ at a certain $Pe$. Regarding the fluctuation of $E_k$, we evaluate the standard deviation of $E_k$ after reaching the statistical steady state in figure \ref{fig:fig8}(c).

The four regimes are as follows:
\begin{itemize}[leftmargin=*,labelindent=6mm,labelsep=3mm]
		\item Regime I ($Pe\leq8\pi$): the system is stable.
		\item Regime II ($8\pi < Pe\leq 16\pi$): the system becomes unstable. Single plumes generate as can be seen in figure \ref{fig:fig8}(b), and thus the dominant wavenumber is $1$.
		\item Regime III ($16\pi <Pe\lesssim 603$): the initial wavenumber $n$ becomes larger than one, and it increases with $Pe$. The trigger of higher-order mode can be explained by the linear stability curve in figure \ref{fig:fig5}. As $Pe>16\pi$, the perturbation of wavenumber $n>1$ becomes unstable. For high enough $Pe$, a higher wavenumber mode can grow even faster than the single wavenumber mode. After a while, the individual plumes merge into a single large one, and the system reaches an asymptotic state with constant $E_k$ ($\sigma=0$ shown in figure \ref{fig:fig8}(c));
		\item Regime IV ($Pe \gtrsim 603$): the plume emission and merging happen continuously even after reaching statistically steady state, and therefore $E_k$ fluctuates with time ($\sigma>0$).
\end{itemize}

Figures \ref{fig:fig8}(a) and (b) indicate that the exponential growth rate and the number of plumes generated initially can be approximately predicted by linear stability analysis. However, at high $Pe$, there is a small deviation between the theory and our simulation.  An explanation is that at high $Pe$, various wavenumbers are excited simutaneously, such that the average growth rate becomes lower than the maximum growth rate predicted by linear stability analysis (equation (\ref{eqjenvelop})).

\subsection{Dependence on Schmidt number}
\begin{figure}
\centering \includegraphics[width=0.8\textwidth]{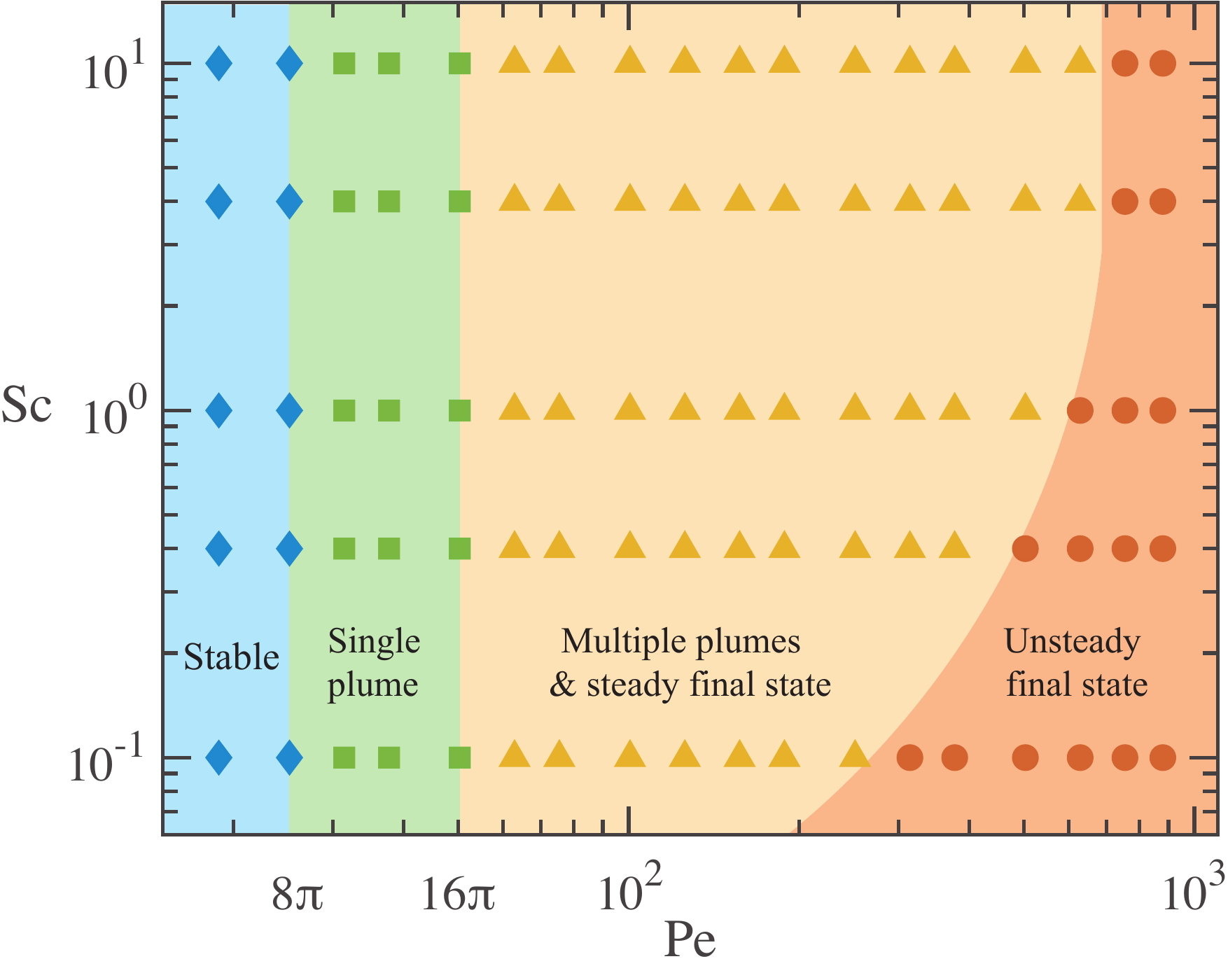}
\caption{\label{fig:fig9} The phase diagram for the case of the catalytic plane with different $Sc$ and $Pe$: For $Pe<8\pi$, the system is stable; for $8\pi<Pe<16\pi$, the system becomes unstable and a single plume is generated; finally, for $Pe>16\pi$, multiple plumes are generated. For the last regime, there are two sub-regimes: for low $Pe$ (orange triangle), multiple plumes eventually merge to a single one and for higher $Pe$ (red circle), there is a newly found regime where the smaller plumes are continuously regenerated. The underlay colors are to guide the eyes.}
\end{figure}

Based on the same classification criteria, we work out the full phase diagram in the $(Pe,$ $Sc)$ parameter space, for $0.1\le Sc \le 10$. Figure \ref{fig:fig9} shows the four different regimes, namely the stable regime (I), the single plume regime (II), the multiple plume regime with a steady final state (III), and the regime with an unstable final state (IV). The transition points between the stable and the unstable regime ($Pe=8\pi$), and between the single plume and the multiple plume regime ($Pe=16\pi$) are insensitive to $Sc$. This can be understood from the linear stability analysis where the onset $s$ for the $n$th wavenumber is $Pe=8\pi n$, independent of $Sc$. However, the onset of regime IV occurs at smaller $Pe$,  provided $Sc<1$. When $Sc \ge 1$, the onset $Pe$ of regime IV becomes independent of $Sc$. 

\begin{figure}
\centering \includegraphics[width=0.95\textwidth]{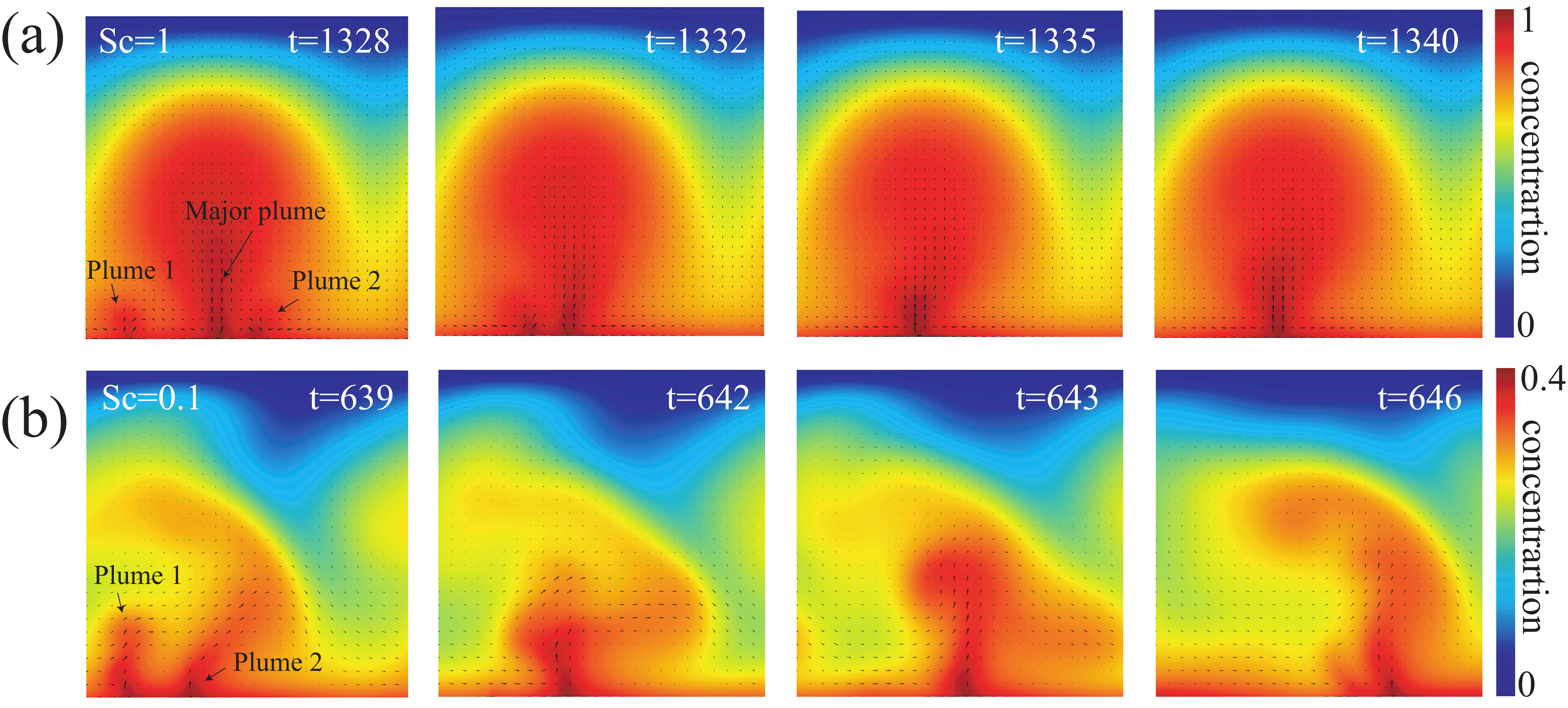}
\caption{\label{fig:Sc01ustd} The concentration contours of plume emission and merging for (a) $Sc=1$ and (b) $Sc=0.1$ with $Pe=754$. For $Sc=1$, the small plumes merge into the stable major plume, and fluctuations are limited in the near-boundary region, while for $Sc=0.1$, the plume merging causes strong fluctuations in the bulk.}
\end{figure}

To further understand why the onset of regime IV behaves differently for $Sc<1$ and $Sc\ge 1$, we have a close inspection on the event of the plume emission and merging for $Sc = 0.1$ and $Sc = 1$ shown in figure \ref{fig:Sc01ustd}. First, for both cases when Pe is large enough, chaotic plume emissions are observed near the catalytic surface. However, the dynamics of the concentration plume are different for large and small $Sc$: For $Sc = 1$ as shown in figure \ref{fig:Sc01ustd} (a), the emitted small plumes gradually merge into the domain-sized plume, and this large plume is relatively stable. Thus, the velocity and concentration fluctuations are limited to near the vicinity of the catalytic surface without penetrating into the bulk region. In contrast, for $Sc = 0.1$ as shown in figure \ref{fig:Sc01ustd} (b), separate plumes merge and eventually be energetic enough to penetrate into the bulk, causing strong fluctuations in the bulk region.

\begin{figure}
\centering \includegraphics[width=0.6\textwidth]{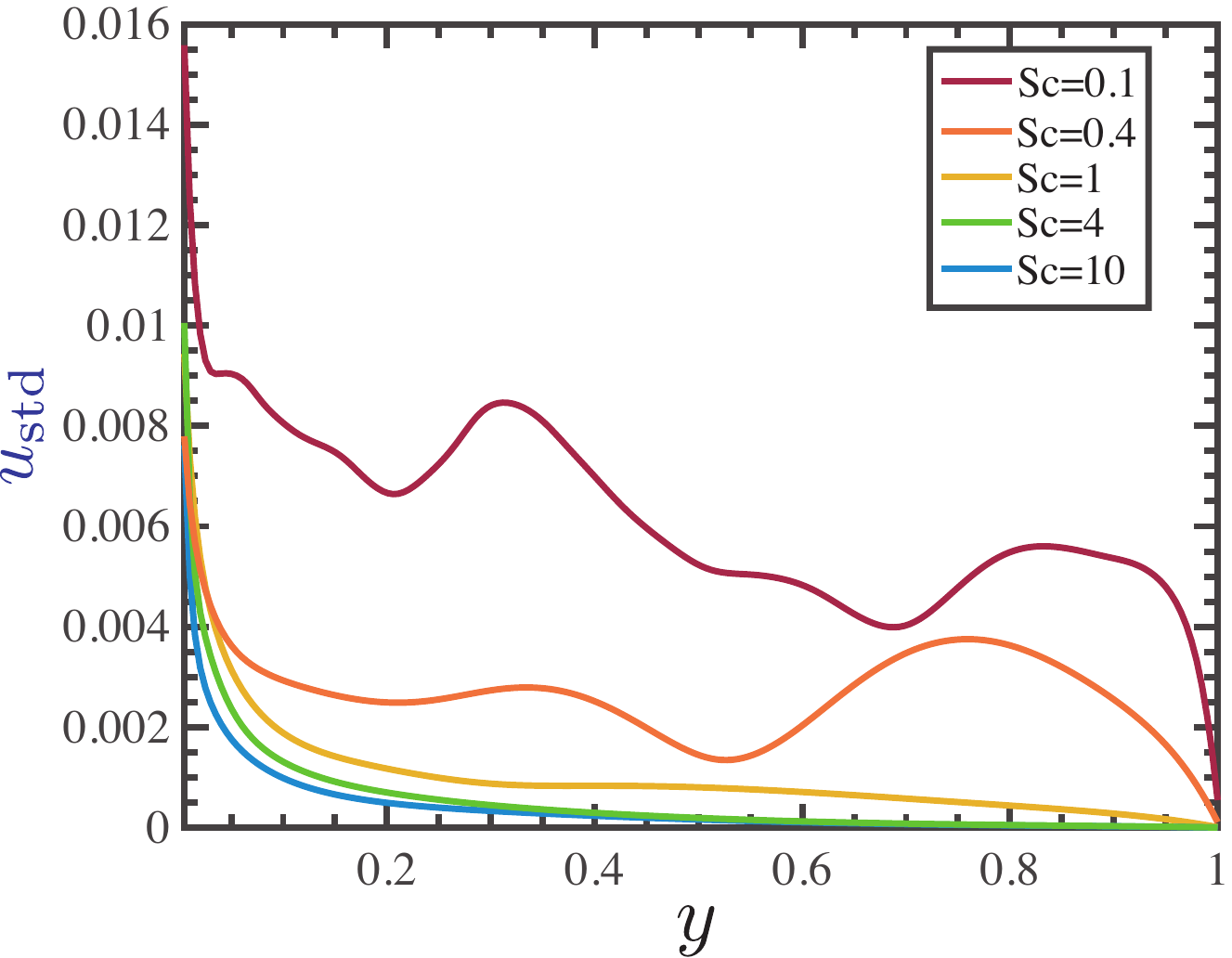}
 \caption{\label{fig:ustd} The standard deviation $u_{std} (y)$ as function of the wall distance $y$ for different Schmidt numbers with P\'eclet number $Pe=754$. Averaging was done of time and over the $x$-direction. All cases belong to regime IV.}
\end{figure}

To quantify this effect, we compute the fluctuation strength, once the system has reached the statistically steady state. It is characterized by the standard deviation of the horizontally averaged horizontal velocity $u_{std} (y)$:
\begin{equation}
\label{eqjustd}
 u_{std}(y)=\langle \sqrt{\langle(u(x,y,t)-\langle u(x,y,t)\rangle _t)^2\rangle _t} \rangle_x,
\end{equation}
where $u(x,y,t)$ is the instantaneous horizontal velocity and $\langle  \rangle $ represents the average over time or $x$-direction, which is denoted by the subscript. The result is plotted in figure \ref{fig:ustd}. From the figure, we find that the fluctuation is maximum at the bottom wall $y=0$ since the diffusiophoretic flow at the wall drives the fluid flow. Moreover, for $Sc<1$, the strong velocity fluctuations are not limited to the near-wall region, but also penetrate into the bulk. In contrast, for $Sc \ge 1$, there are only large fluctuations in the near-wall region and $u_{std}(y)$ is monotonically decaying with wall distance $y$. 

We now understand that the chaotic fluctuations observed in regime IV originate from different physical mechanisms for small and large $Sc$. For small $Sc$, as the fluctuations are mainly contributed from the bulk, one expects that the bulk viscous dissipation plays a role, and thus lower onset $Pe$ should be obtained for smaller $Sc$. However, it does not hold for the situation of large $Sc$ since the fluctuations are mainly contributed by the chaotic plume emission close to the catalytic surface. To work out the details of the chaotic plume emission, non-linear stability is worthy to be conducted in the future.



As already mentioned in the introduction, our results share some similarities with those of the B\'enard-Marangoni instability. For both cases, if the P\'eclet or Marangoni number is above a critical value,  the system becomes unstable. \cite{bergeon1998marangoni} comprehensively  studied the Marangoni convection and found that as the Marangoni number increases, the plume will develop into single-roll or multiple-roll structures, which is similar to the single wavenumber or higher-order wavenumber modes observed in regime II and III, respectively.

As a final remark, \cite{michelin2020spontaneous} have shown the diffusio-phoretic instability in a confined phoretic channel, from which they also observe the generation of the plumes. Note that \cite{michelin2020spontaneous} have also considered the non-linear terms in the advection-diffusion equation, however, for the momentum equation, they consider the case of Stokes flow, such that the non-linear terms and effects of Schmidt number have not been considered. The chaotic plume emission observed in regime IV is the unique feature for high P\'eclet numbers, which, however, has not been focused on in most of the previous studies. Moreover, with an analytical calculation, we obtain the dominant wavenumber and its growth rate for different $Sc$ and $Pe$, which agrees with our simulation.

\section{Simulation of the phoretic particle}\label{particle} 
Given the analysis of the catalytic plane, we now conduct three-dimensional simulations of a spherical phoretic particle to study the effect of plume emission and merging on the particle motion. 
\subsection{Numerical setup}
The set-up is as follows: a phoretic particle is positioned at the center of the domain, and then due to diffusio-phoresis, the particle will self-propel.  The governing equations consist of two parts. The first is the same as that in section \ref{nummethod}, which is to solve the three-dimensional version of equations (\ref{eqj2}) and (\ref{eqj4}), except for the characteristic length which now is the radius of the particle. The second part involves the governing equation for the dynamics of the phoretic particle. However, one faces the challenge of dealing with a moving immersed boundary condition. To deal with it, we make use of moving least squares (MLS) based immersed boundary (IB) method, where the particle interface is represented by a triangulated Lagrangian mesh. For details of our MLS-based IB method, we refer to \cite{spandan2017parallel}. The concentration boundary condition is that the wall normal concentration gradient is a constant $ \frac{\partial c}{\partial n} =- 1$, which can be achieved by forcing the concentration at the particle surface based on the concentration interpolated at the probe located at a short distance (1 grid size) from the surface of the particle. The velocity boundary condition is 
\begin{equation}
\label{eqj15_1}
u_s=\nabla_s c,
\end{equation}
where $u_s$ is the surface gradient ($\nabla_s$) of the concentration. 

\begin{figure}
\centering \includegraphics[width=0.9\textwidth]{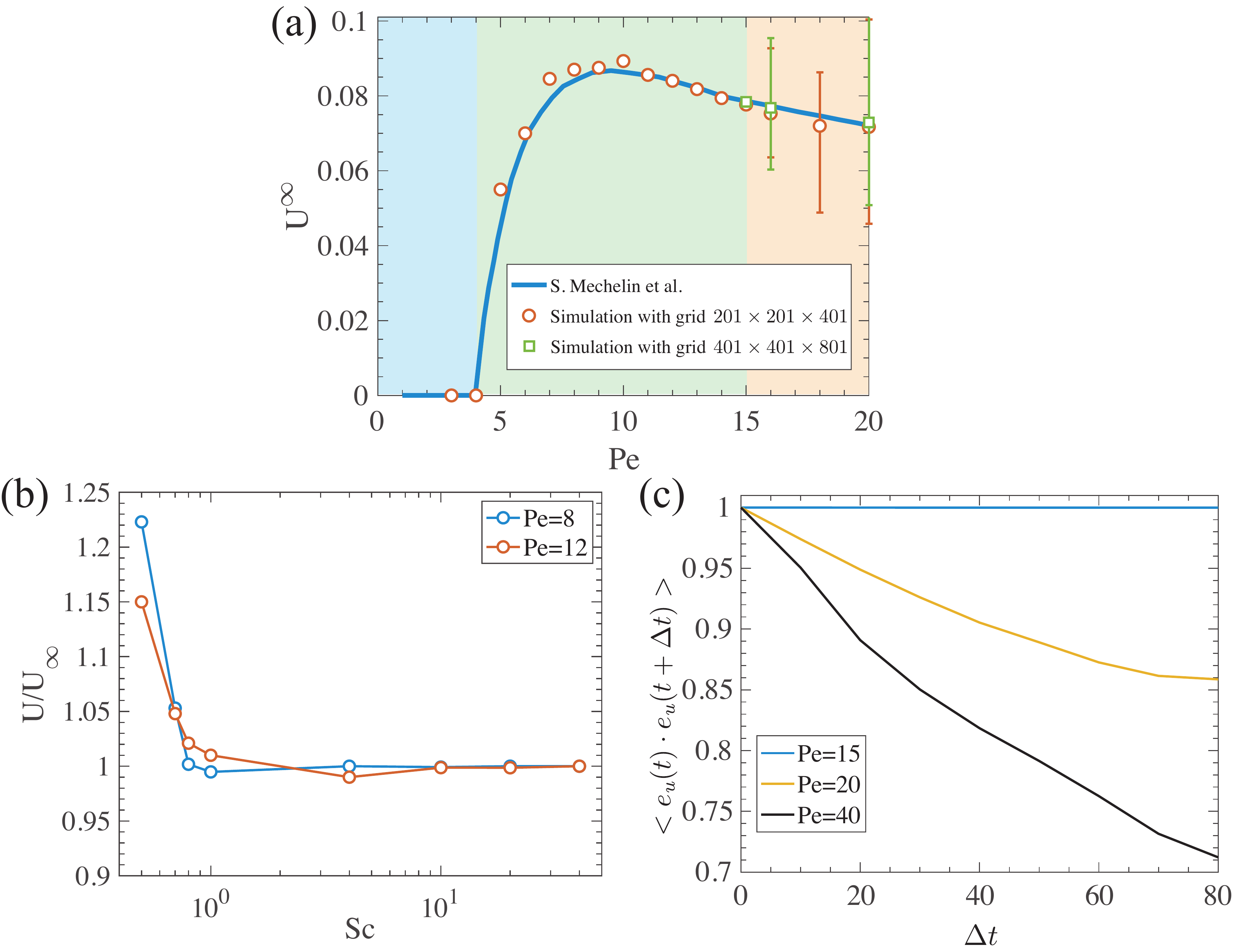}
\caption{\label{fig:fig11} (a) The terminal velocity $U^\infty$ of phoretic particles as function of $Pe$ for $Sc=1$. The result from the axisymmetric simulation by \cite{michelin2013spontaneous} is shown as blue solid curve. Our results for the full three-dimensional case of different grid size are indicated by red circles (grid $201 \times 201 \times 401$) and green squares (grid $401 \times 401 \times 801$). The points for $Pe>15$ indicate the average terminal velocity and the range of fluctuations is shown by the solid bars. The motion of the phoretic particle is divided into three different regimes: stable, symmetry breaking, and chaotic motion due to plume generation. (b) The nomalized terminal velocity of different $Sc$ for the case $Pe=8$ and $12$. The velocity is normalized by the terminal velocity at $Sc=40$. The result shows that when $Sc>1$, the terminal velocity converges to a constant.} (c) The temporal auto-correlation function of the unit direction vector for three different $Pe$ for $Sc=1$. The temporal auto-correlation indicates whether the particle performs chaotic motion or not. 
\end{figure}

The domain size is $L_x \times L_y \times L_z=20R \times 20R \times 40R$, in terms of the particle radius $R$. We use uniform grids $N_x \times N_y \times N_z =201 \times 201 \times 401$. Mesh refinement tests are done at $Pe=15,16$ and $20$ with doubled grid numbers in each dimension. Figure \ref{fig:fig11}(a) indicates that the result for the grid $401 \times 401 \times 801$ is nearly indistinguishable from that for $201 \times 201 \times 401$.

For the spherical particle, the radius $R$ is used as the length scale in P\'eclet number:
\begin{equation}
\label{eqjPe_par}
Pe=\frac{M \alpha R}{D^2}. \quad
\end{equation}
We will present the result of phoretic particles for different $Pe$ from 3 to 20 with $Sc=1$. 

\begin{figure}
\centering
\centering \includegraphics[width=0.9\textwidth]{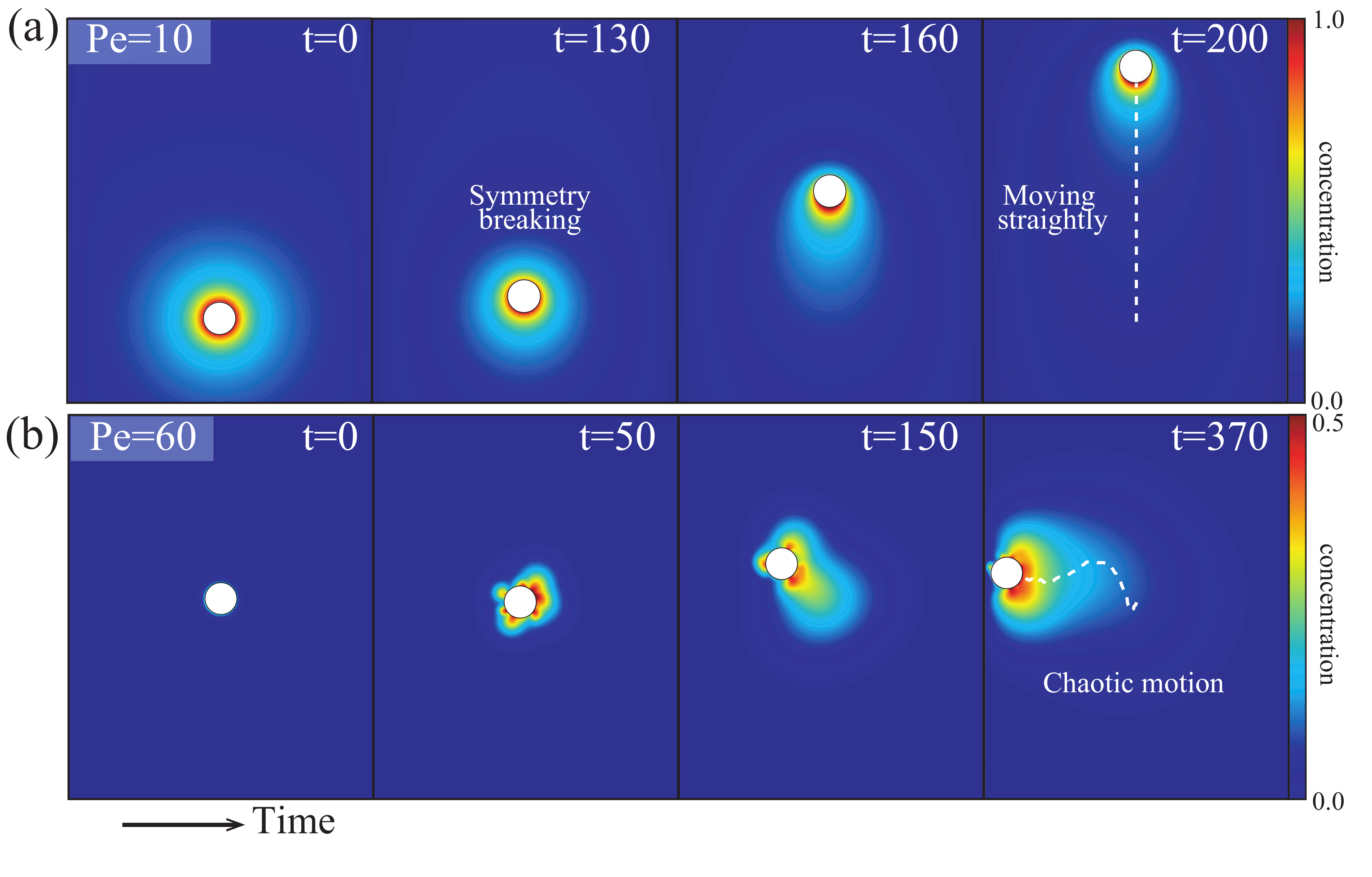}
\caption{\label{fig:fig10}The concentration cross-section from three-dimensional simulations of an isotropic catalytic particle for $Pe=10$ and $60$. Again, $Sc=1$. The simulation is at a domain $L_x \times L_y \times L_z = 20R\times20R\times40R$, in terms of the particle radius $R$. The grids are $201 \times 201 \times 401$. To better demonstrate the chaotic trajectory, the motion of the particle in (b) is projected to x-z plane. (a) $Pe=10$, the particle moves straightly. (b) $Pe=60$, plumes are generated at the surface of the particle, which starts to move irregularly.}
\end{figure}

\subsection{Result of the phoretic particle} \label{res_par}
Similar to the case of the catalytic plane,  the diffusio-phoretic instability breaks the rotational-symmetry of the phoretic particle.  It has been shown by \cite{michelin2013spontaneous} that the phoretic particle breaks the symmetry when $Pe$ is larger than 4. Therefore, as a validation, we first simulate cases with small $Pe$ and compare to the results obtained from \cite{michelin2013spontaneous}. In figure \ref{fig:fig11}(a), we plot the numerical terminal velocity $U^{\infty}$. For $Pe>4$, indeed symmetry breaking occurs (e.g. figure \ref{fig:fig10} (a) for $Pe=10$) and the particle moves along a straight trajectory. The terminal velocities agree with those obtained from \cite{michelin2013spontaneous}. Furthermore, in figure \ref{fig:fig11}(b) we check whether the terminal velocity is sensitive to the Schmidt number $Sc$. Interestingly, the figure suggests that when $Sc\geq1$ for $Pe=8$ and $12$, the terminal velocity converges to a constant. However,  to understand why terminal velocity levels off, further study is needed in the future. Regarding the grid resolution requirement, the necessary grid resolution will increase dramatically for very large Sc. In order to run as many cases as possible to fully explore the phase diagram, we stick to $Sc=1$ for the rest of our simulations.

After this validation we now extend the calculations to higher $Pe$. Multiple plumes emission and merging occur at the surface of the phoretic particle (e.g. figure \ref{fig:fig10} (b)  for $Pe=60$), which is similar to that observed for the catalytic plane. The continuously emitted plumes change the direction of the phoretic particle and lead to chaotic motion. 

To characterize the motion of the particle, we calculate the mean temporal auto-correlation of the particle direction:
\begin{equation}
\label{eqj15_2}
\langle\pe_u(t) \cdot \pe_u(t+\Delta t)\rangle=\frac{1}{T}\int_{0}^{T} \pe_u(t) \cdot \pe_u(t+\Delta t)dt, \quad
\end{equation}
where $\pe_u(t)=\pu(t)/|\pu(t)|$ is the unit direction vector of the particle velocity at $t$. The integral upper limit $T$ is chosen large enough to achieve statistical stationary. The auto-correlation for different $Pe$ is shown in figure \ref{fig:fig11} (c). When $Pe=20$, the auto-correlation becomes considerably less than 1 with increasing $\Delta t$, which means that the particle starts to meander in different directions. The chaotic behavior of the particle has also been shown by the fluctuation of velocity in figure \ref{fig:fig11}(a), which is represented by the solid bars. Interestingly, for $Pe \gtrsim 15$, the average velocity (the red circle) still lies near the result by \cite{michelin2013spontaneous}, but for larger $Pe$, the velocity shows larger fluctuation, i.e. more chaotic behavior. 

Thus we can classify the motion for a phoretic particle into three regimes: 
	\begin{itemize}[leftmargin=*,labelindent=6mm,labelsep=3mm]
		\item $Pe<4$, the particle remains stable.
		\item $4<Pe\lesssim15$, symmetry breaking occurs and the particle moves straight.
		\item $Pe\gtrsim15$, the particle moves chaotically.
	\end{itemize}

Figure \ref{fig:fig10} (b) shows that the plumes are continuously generated and merge, which alters the concentration distribution and steer the moving direction of the phoretic particle. This shows that our newly found regime IV in figure \ref{fig:fig9} leads to the chaotic motion for the case of phoretic particles. 

\subsection{Comparison between phoretic particle and catalytic plane}
We now compare the various regimes for the catalytic plane (Section \ref{sim}) with those for the phoretic particle (Section \ref{particle}). The similarity between the two setups is that both the instability and chaotic flow can be observed for both setups. A major difference between them is that the regimes for the catalytic plane are classified by the distinct plume dynamics whereas the regimes for the phoretic particle are classified by the distinct particle motions. However, both regimes II and III for the catalytic plane lead to a steady final state with a single plume, which for the case of the phoretic particle implies straight motion. For classification of the particle motion, the same fate of particle motion leads one to define only one regime despite the distinct plume dynamics during the initial transient stage.

Another difference between the catalytic plane and phoretic particle is that the onset P\'eclet numbers are different. However, this difference simply reflects that the characteristic length scales in the definition of $Pe$ are different for both systems.

We note that also \cite{PhysRevLettHu} numerically observed the chaotic motion of phoretic particles. However, the plume generation and merging, which could provide a route to understand the chaotic motion of the particle at high enough $Pe$, were not studied in that paper. Besides, in the experiments of active droplets, it is also observed that the droplet can move in a helical or even chaotic trajectory at high $Pe$ \citep{suga2018self, maass2016swimming}. \cite{morozov2019orientational} also observed the helical and chaotic motion of the catalytic particle.Recently, the stochastic dynamics of active particles was analysed \citep{gaspard2018fluctuating,chamolly2019stochastic}. Based on the stochastic approach using Langevin equations, the active particle motion is split into a diffusive part and a ballistic part. In our work here, it is the deterministic plume emission that is the source of the diffusive motion, and an one-to-one comparison is difficult due to the quite different natures of the approaches. Very recently, \cite{babak2020} observed plume generation and merging at the surface of a meandering chemically active droplet. This recent finding reflects the importance of the plume dynamics in determining the droplet motion. Here, for the diffusio-phoretic particle, we have also revealed such plume generation and merging phenomena and have related it to the instability of the flow near the surface.

\section{Concluding remarks} \label{sec:conc} 

In summary, we have studied the instability driven by diffusio-phoretic effects at the interfaces for two different systems: a catalytic plane and a spherical phoretic particle. The P\'eclet number ($Pe$) and Schmidt number ($Sc$) are the parameters that determine the states of the system. 

For a catalytic plane, via linear stability analysis, we quantitatively studied the growth of various wavenumber perturbation. With assistance of the simulation, we have classified four regimes for different $Pe$ and $Sc$ based on the exponential growth rate of the instability, number of plumes generated initially and fluctuation of the kinetic energy after reaching the statistically steady state ($E_k$). For $Pe \leq 8\pi$, the system is stable. For $8\pi < Pe \leq 16\pi$, the system becomes unstable, a single plume is generated and the system reaches a steady state eventually. For $Pe>16\pi$, multiple plumes are generated initially, which merge into a single one to attain a stable state eventually due to non-linear saturation. However, for even higher $Pe$ ($Pe\gtrsim603$ for $Sc=1$), small plumes are continuously generated and merge with each other, the system remains unstable and therefore $E_k$ fluctuates in time.

Based on the linear stability analysis, we understand that the onset $Pe$ between regime I and II, and regime II and III are independent of $Sc$. However, there is noticeable effect of $Sc$ on the transition to regime IV, which is associated with different flow structures for different $Sc$. For small $Sc$, the strong fluctuations of concentration and kinetic energy also occur in the bulk region, whereas for large $Sc$, the fluctuations are only contributed by the chaotic plume emission close to the catalytic plane. As the viscous dissipation in the bulk plays a role for small $Sc$ cases, lower onset P\'eclet numbers of regime IV are obtained for lower $Sc$. However, it does not hold for large $Sc$.



Then we extend our research to three-dimensional simulations of the spherical phoretic particle. Despite the geometric difference, an analogous phenomenon happens at the surface of the particle which triggers different particle motions. Similar to the case of the catalytic plane, for the case at $Sc=1$,  when $Pe>4$, the particle starts to break the symmetry. For higher $Pe\gtrsim15$, also similar to the observation of the catalytic plane, the small plumes start to be generated continuously at the surface of the particle, which will steer the particle and lead to meandering motion. The analogous phenomenon indicates that the chaotic motion of the phoretic particle results from the instability at the interface driven by diffusio-phoretic effects.

The present work makes a contribution to the understanding of the diffusio-phoretic instability. First, the study reveals the existence of a highly unstable regime at high $Pe$. We not only study the onset $Pe$ of the unstable mode, but also analytically work out the dominant wavenumber as a function of $Pe$. For high enough $Pe$ (regime III), multiple plumes are emitted into the surrounding fluid. For even higher $Pe$ (regime IV), smaller wavelength perturbations are dominant, which leads to continuous plume generation and subsequently to chaotic flow. Second, our results show that the diffusio-phoretic instability at the catalytic surface can eventually lead to chaotic motion of the phoretic particle. Through simulations of the phoretic particle at high $Pe$, we not only see its chaotic motion, but also observe the plume emission and merging events near the surface of the particle, which is similar to the situation of regime IV for the case of the catalytic plane. The study of the phoretic plane thus provides a framework to understand the motion of the phoretic particle.

Many questions remain open. For example, how does the particle motion and flow field change for phoretic particles in a complicated environment, such as phoretic particle near a wall? How does the plume generation and merging change the collective behavior of  phoretic particles? How about the effect of plume generation on rod particles rather than spheres? Building on the here obtained insight into the mechanism behind the chaotic motion of phoretic particles, it is worthwhile to further explore the effects of the plume generation on the motion of particles in the more complicated setups as mentioned above, in particular, on collective effects.

\section*{Acknowledgements}
We greatly appreciate the valuable discussions with Maziyar Jalaal, Chong Shen Ng, Qi Wang, Babak Vajdi Hokmabad, Corinna Maass and Andrea Prosperetti. We acknowledge the support from the Netherlands Center for Multiscale Catalytic Energy Conversion (MCEC), an NWO Gravitation program funded by the Ministry of Education and support from the ERC-Advanced Grant "DDD" under the project number 740479. We acknowledge that the results of this research have been achieved using the DECI resource Kay based in Ireland at Irish HPC center with support from the PRACE. We also acknowledge PRACE for awarding us access to MareNostrum at the Barcelona Supercomputing Centre (BSC) under PRACE project number 2017174146 and JUWELS at the J\"ulich Supercomputing Centre. This work was also partly carried out on the national infrastructure of SURFsara with the support of SURF Cooperative, the collaborative ICT organization for Dutch education and research.

\section*{Declaration of interests}
The authors report no conflict of interest.

\appendix
\section{Linear stability analysis for catalytic plane} \label{Aappendix}
In this appendix, the linear stability analysis is performed to investigate the stability of the system (\ref{eqj2})-(\ref{eqj8}).

\subsection{Base flow}
The base flow can be obtained by assuming a static flow, 
\beq \lb{eq.u0}
\bar{\pu} = \mathbf{0}.
\eeq 
Substituting equation \er{eq.u0} into \er{eqj4} we obtain the pressure solution, 
\beq \lb{eq.p0}
\bar{p} = 0.
\eeq 
Substituting equation \er{eq.u0} into \er{eqj2}, we obtain
\beq \lb{eq.c0-1}
\fr{\pat \bar{c}}{\pat t} = \fr{1}{Pe} \fr{\pat^2 \bar{c}}{\pat y^2}, \quad 
\left. \fr{\pat \bar{c}}{\pat y} \right|_{y=0} = -1, \quad 
\left. \bar{c} \right|_{t=0} = 0.
\eeq 
Denote the Laplace transform as \citep{liu2017unified}
\beq 
 \bar{C} = \cL( \bar{c})  = \int_0^\infty \bar{c} e^{-\alpha t} dt, \quad 
\bar{c} = \cL^{-1} ( \bar{C}) = \fr{1}{2\pi i}\int_{-i \infty}^{i \infty}  \bar{C} e^{\alpha t} dt.
\eeq 
Note that
\beq 
\cL  \left( \fr{\pat \bar{c}}{\pat t} \right) = \alpha  \bar{C}, \quad \cL \left( 1 \right) = \fr{1}{\alpha}.
\eeq 
Then \er{eq.c0-1} can be transformed to
\beq \lb{eq.c0-2}
\fr{d^2  \bar{C}}{d y^2} - Pe \alpha \bar{C} =0, \quad 
\left. \fr{d  \bar{C} }{d y} \right|_{y=0} = -\fr{1}{\alpha}.
\eeq 
The solution of \er{eq.c0-2} that vanishes at infinity is 
\beq \lb{eq.c0-laplace}
\bar{c} = \fr{1}{\alpha} \cdot \fr{1}{ \sqrt{Pe \alpha} } e^{- y \sqrt{Pe \alpha} }.
\eeq 
Recall that (\cite{jeffrey2007table}, p.~1110)
\beq 
\cL^{-1} \left( \fr{1}{ \alpha} \right) = 1, \quad 
\cL^{-1} \left( \fr{1}{ \sqrt{Pe  \alpha} } e^{- y \sqrt{Pe  \alpha} } \right) = \frac{ 1 }{\sqrt{\pi Pe t}} e^{ - \frac{Pe y^2}{4 t} },
\eeq 
and use the convolution theorem. Then the concentration solution in physical space can be obtained as (\cite{Wu2006}, p. 144), 
\beq \lb{eq.c0}
\bar{c} = \int_0^t \frac{ 1 }{\sqrt{\pi Pe (t-\tau)}} \exp \left[ - \frac{Pe y^2}{4 (t-\tau)} \right] \textrm{d} \tau.
\eeq
The corresponding derivative is 
\beq \lb{eq.dc0-1}
\frac{\pat \bar{c} }{\pat y} = - \int_{ \xi_0 }^\infty \frac{ 2 }{\sqrt{\pi } } e^{ - \xi^2 } \textrm{d} \xi, \quad 
\text{with } \xi^2 = \frac{ Pe y^2  }{ 4 (t-\tau)}, \quad 
\xi_0^2 = \frac{ Pe y^2  }{ 4 t }.
\eeq 
Considering the limit $t\rightarrow \infty$, i.e., $\xi_0\rightarrow0$, we obtain
\beq \lb{eq.dc0-1}
\frac{\pat \bar{c}}{\pat y} = -1.
\eeq 

\subsection{Perturbation flow} 	

Substituting the base flow \er{eq.u0}, \er{eq.p0} and \er{eq.dc0-1} into the governing equations \er{eqj2} and \er{eqj4} and keeping only the $O(\epsilon)$-terms, we get the linearized governing equations \er{eqj9} and \er{eqj11}.
The boundary conditions are equations \er{bc12}.

The perturbation is assumed as \er{eqj14}. Substituting the perturbation term (\ref{eqj14}) into the governing equations \er{eqj9} and \er{eqj11}, and re-arranging the resultant equations, we obtain 
	\beqn
	& & \left( \pat_y^2 - k^2 \right)  \check p = 0, \\
	& & \left( \pat_y^2 - k^2 - \frac{s Pe}{Sc} \right) \check v = \pat_y \check p, \lb {eq.w1}\\
	& & \left( \pat_y^2 - k^2 - s Pe \right) \check c =  -  Pe \check v. \lb {eq.c1}
	\eeqn
	The boundary conditions are 
	\beq\lb{bc-1-y}
	\left. \fr{d \check c}{d y} \right|_{y=0} = 0, \quad
	\left. \fr{d \check v}{d y} \right|_{y=0} = k^2 \check c, \quad 
	\left. \check v \right|_{y=0} = 0. 
	\eeq 
	The pressure solution that decays at infinity is 
	\beq \lb{eq.p1-sol}
	\check p = e^{-k y}, \quad k  = 2 \pi n, \quad n\in \textbf{N} .
	\eeq
	Substituting equation \er{eq.p1-sol} into \er{eq.w1}, and using the third equation in \er{bc-1-y}, the vertical velocity is found to be
	\beq\lb{eq.w1-sol}
	\check v = \frac{ k }{s} \left( e^{-k y} - e^{-k y \sqrt{1+ \frac{s Pe}{ Sc k ^2}} } \right).
	\eeq
	Similarly, by substituting equation \er{eq.w1-sol} into \er{eq.c1}, and using the first equation in \er{bc-1-y}, the concentration is
	\beqn \lb{Ac}
	\check c &=&  \frac{k}{s^2} \left( e^{-k y} - \frac{ e^{-k y \sqrt{1+ \frac{s Pe}{ k^2}} }}{ \sqrt{1+ \frac{s Pe}{k^2}} } \right)  \nonumber \\
	& &\   - \frac{k}{s^2} \sqrt{1+ \frac{s Pe}{ Sc k^2}} \frac{Sc }{Sc -1} \left( \frac{ e^{-k y \sqrt{1+ \frac{s Pe}{ Sc k^2}} }}{ \sqrt{1+ \frac{s Pe}{ Sc k^2}} } - \frac{ e^{-k y \sqrt{1+ \frac{s Pe}{ k^2}} }}{ \sqrt{1+ \frac{s Pe}{ k^2}} } \right).
	\eeqn
	It seems that equation \er{Ac} could be invalid for $Sc=1$ since the denominator $Sc-1$ therein is zero. However, this is not the case since the term in the brackets of the second line also reduces to zero. By performing a Taylor series expansion of \er{Ac} at $Sc=1$ (rule of l'Hospital), one can easily prove that
\beq\lb{Alim}
\check c =  \frac{ k }{s^2} \left( e^{-k y} - \frac{ e^{-k y \sqrt{1+ \frac{s Pe}{ k^2}} }}{ \sqrt{1+ \frac{s Pe}{k^2}} } \right)
- \frac{Pe \left( 1+ ky \sqrt{1+\frac{sPe}{k^2}} \right) }{ 2 s k  \left( 1+ \fr{sPe}{k^2} \right) } e^{-ky\sqrt{1+\frac{sPe}{k^2}}}. \nonumber 
\eeq
	Finally, substitute \er{eq.w1-sol} and \er{Ac} into the second equation in \er{bc-1-y}, we can obtain the equation \er{eqj14_5} that determines the exponential growth rate $s=s(k; Pe, Sc)$.

We now explain briefly how to use \er{eqj14_5} to get the theoretical results in Figure 3.\\
Define
\beq
\delta^2-1 =\fr{s Pe}{k^2}.
\eeq
Then
\beqn\lb{eq.s-k-1}
Pe&=& Pe(\delta; k, Sc) = k \delta \left( \delta+1 \right)  \left( \delta+ \sqrt{1+  \fr{\delta^2-1}{Sc}  }  \right), \\
s &=& s(\delta; k, Sc) = \fr{k (\delta^2-1)}{ \delta \left( \delta+1 \right)  \left( \delta+ \sqrt{1+  \fr{\delta^2-1}{Sc}  }  \right) }. \lb{eq.s-k-2}
\eeqn
Thus, for given $k$ and $Sc$, we obtain the curve $s$ vs $Pe$ in Figure 3(b) by varying $\delta$. 
Similarly, for given $k$, we obtain the contour $s$ in the $(Pe, Sc)$ plane in Figure 3(a) by varying $\delta$ and $Sc$.

\subsection{Determination of the dominant wavenumber} 

The maximum growth rate, as well as the dominant wavenumber, can also be determined from \er{eqj14_5} or equivalently \er{eq.s-k-1} and \er{eq.s-k-2}. To show this, we rewrite \er{eq.s-k-1} and \er{eq.s-k-2} as
\beqn\lb{eq.s-k-3}
\kappa &=& \ka(\delta; Pe, Sc) = \fr{1} { \delta \left( \delta+1 \right)  \left( \delta+ \sqrt{1+  \fr{\delta^2-1}{Sc}  } \right) }, \\
\sigma &=& \sigma(\delta; Pe, Sc) = \fr{\delta^2-1} { \left[ \delta \left( \delta+1 \right)  \left( \delta+ \sqrt{1+  \fr{\delta^2-1}{Sc}  }  \right) \right]^2}, \lb{eq.s-k-4}
\eeqn
where
\beq 
\ka = \fr{k}{Pe}, \quad 
\sigma = \fr{s}{Pe}. 
\eeq
For certain Pe and Sc, the maximum growth rate is obtained by $\partial_k s=0$, which is equivalent to 
\beq \lb{eq.s-d}
\fr{d\sigma}{d\ka} = \fr{d\sigma}{d\delta} \left( \fr{d\kappa}{d\delta} \right)^{-1} = 0,
\quad  \text{implying} \quad 
\fr{d\sigma}{d\delta}=0.
\eeq 
By solving \er{eq.s-d} and denoting the solution as $\delta_e$, the maximum growth rate and the corresponding dominant wavenumber are always given by 
\beq 
s = \sigma_e Pe, \quad 
k = \ka_e Pe,
\eeq 
where $\sigma_e=\sigma(\delta_e)$ and $\kappa_e=\kappa(\delta_e)$ can be obtained by substituting $\delta_e$ into \er{eq.s-k-4} and \er{eq.s-k-3}, respectively. 
Note that in general equation \er{eq.s-d} needs to be solved numerically. However, for some specific $Sc$, it can also be solved analytically. Two examples are as follows:

\textbf{Example 1:} When $Sc=1$, equations \er{eq.s-k-3}, \er{eq.s-k-4} and \er{eq.s-d} reduce to
\beq\lb{eq.k-s-4}
\kappa = \fr{1} { 2 \delta^2 \left( \delta+1 \right)}, \quad 
\sigma = \fr{\delta -1} { 4 \delta^4 \left( \delta+1 \right)}, \quad
\fr{\textrm{d} \sigma}{\textrm{d} \delta} = \frac{2+\delta-2 \delta ^2 }{2 \delta ^5 (\delta +1)^2} = 0.
\eeq 
The corresponding solution is 
\beq 
\delta_e = \frac{1 + \sqrt{17} }{4}, \quad 
\ka_e = \frac{ 31 - 7 \sqrt{17} }{16}, \quad 
\sigma_e = \frac{85 \sqrt{17} - 349}{128}. 
\eeq

\textbf{Example 2:} When $Sc=\infty$, equations \er{eq.s-k-3}, \er{eq.s-k-4} and \er{eq.s-d} reduce to
\beq\lb{eq.k-s-4}
\kappa = \fr{1} { \delta \left( \delta+1 \right)^2}, \quad 
\sigma = \fr{\delta -1} { \delta^2 \left( \delta+1 \right)^3}, \quad
\fr{\textrm{d} \sigma}{\textrm{d} \delta} = \frac{2 + 4 \delta-4 \delta^2 }{\delta ^3 (\delta +1)^4} = 0.
\eeq 
The corresponding solution is 
\beq \lb{infty}
\delta_e = \frac{1 + \sqrt{3} }{2}, \quad 
\ka_e = 2 \sqrt{3} - \frac{10}{3}, \quad 
\sigma_e = \frac{4}{9} \left( 26 \sqrt{3} - 45 \right).
\eeq

\bibliographystyle{jfm}
\bibliography{reference_catalyticplane}

\end{document}

%% file: UDSymbol.tex
\newcommand{\bsub}{\begin{subequations}}
\newcommand{\esub}{\end{subequations}}

\newcommand{\ka}{\kappa}

\newcommand{\cL}{\mathcal L}

\newcommand{\pe}{\textbf{\emph{e}}}

\newcommand{\pu}{\textbf{\emph{u}}}

\newcommand{\pat}{\partial}

\newcommand{\beq}{\begin{equation}}
\newcommand{\eeq}{\end{equation}}
\newcommand{\bsubeq}{\begin{subequations}}
\newcommand{\esubeq}{\end{subequations}}
\newcommand{\beqn}{\begin{eqnarray}}
\newcommand{\eeqn}{\end{eqnarray}}
\newcommand{\fr}{\frac}
\newcommand{\lb}{\label}
\newcommand{\er}{\eqref}

\graphicspath{{figures/}}

%% file: manuscript.bbl
\begin{thebibliography}{48}
\expandafter\ifx\csname natexlab\endcsname\relax\def\natexlab#1{#1}\fi
\def\au#1{#1} \def\ed#1{#1} \def\yr#1{#1}\def\at#1{#1}\def\jt#1{\textit{#1}}
  \def\bt#1{#1}\def\bvol#1{\textbf{#1}} \def\vol#1{#1} \def\pg#1{#1}
  \def\publ#1{#1}\def\arxiv#1{#1}\def\org#1{#1}\def\st#1{\textit{#1}}

\bibitem[Anderson(1989)]{anderson1989colloid}
{\sc \au{Anderson, J.~L.}} \yr{1989}  \at{Colloid transport by interfacial
  forces}.  \jt{Annu. Rev. Fluid Mech.}  \bvol{21}~(1),  \pg{61--99}.

\bibitem[B{\"a}r {\em et~al.\/}(2020)B{\"a}r, Gro{\ss}mann, Heidenreich \&
  Peruani]{bar2019self}
{\sc \au{B{\"a}r, M.}, \au{Gro{\ss}mann, R.}, \au{Heidenreich, S.} \&
  \au{Peruani, F.}} \yr{2020}  \at{Self-propelled rods: Insights and
  perspectives for active matter}.  \jt{Annu. Rev. Condens. Matter Phys.}
  \bvol{11},  \pg{441--466}.

\bibitem[Bergeon {\em et~al.\/}(1998)Bergeon, Henry, Benhadid \&
  Tuckerman]{bergeon1998marangoni}
{\sc \au{Bergeon, A.}, \au{Henry, D.}, \au{Benhadid, H.} \& \au{Tuckerman,
  L.~S.}} \yr{1998}  \at{Marangoni convection in binary mixtures with soret
  effect}.  \jt{J. Fluid Mech.}  \bvol{375},  \pg{143--177}.

\bibitem[Boeck \& Vitanov(2002)]{boeck2002low}
{\sc \au{Boeck, T.} \& \au{Vitanov, N.~K.}} \yr{2002}  \at{Low-dimensional
  chaos in zero-prandtl-number benard--marangoni convection}.  \jt{Phys. Rev.
  E}  \bvol{65}~(3),  \pg{037203}.

\bibitem[Bray(2000)]{bray2000cell}
{\sc \au{Bray, D.}} \yr{2000} {\em Cell movements: from molecules to
  motility\/}.  \publ{Garland Science}.

\bibitem[Chamolly \& Lauga(2019)]{chamolly2019stochastic}
{\sc \au{Chamolly, A.} \& \au{Lauga, E.}} \yr{2019}  \at{Stochastic dynamics of
  dissolving active particles}.  \jt{Eur. Phys. J. E}  \bvol{42}~(7),  \pg{88}.

\bibitem[Davis(1987)]{davis1987thermocapillary}
{\sc \au{Davis, S.~H.}} \yr{1987}  \at{Thermocapillary instabilities.}  \bt{In
  {\em Annu. Rev. Fluid Mech.\/}},  \pg{pp. 403--435}.  \publ{Annual Reviews
  Inc}.

\bibitem[Drazin \& Reid(2004)]{drazin2004}
{\sc \au{Drazin, P.~G.} \& \au{Reid, W.~H.}} \yr{2004} {\em Hydrodynamic
  Stability\/}.  \publ{Cambridge: Cambridge University Press}.

\bibitem[Ebbens \& Howse(2010)]{ebbens2010pursuit}
{\sc \au{Ebbens, S.~J.} \& \au{Howse, J.~R.}} \yr{2010}  \at{In pursuit of
  propulsion at the nanoscale}.  \jt{Soft Matter}  \bvol{6}~(4),
  \pg{726--738}.

\bibitem[Gaspard \& Kapral(2018)]{gaspard2018fluctuating}
{\sc \au{Gaspard, P.} \& \au{Kapral, R.}} \yr{2018}  \at{Fluctuating
  chemohydrodynamics and the stochastic motion of self-diffusiophoretic
  particles}.  \jt{J. Chem. Phys.}  \bvol{148}~(13),  \pg{134104}.

\bibitem[Golestanian {\em et~al.\/}(2007)Golestanian, Liverpool \&
  Ajdari]{gol2007}
{\sc \au{Golestanian, R.}, \au{Liverpool, T.~B.} \& \au{Ajdari, A.}} \yr{2007}
  \at{Designing phoretic micro-and nano-swimmers}.  \jt{New J. Phys.}
  \bvol{9}~(5),  \pg{126}.

\bibitem[Gradshteyn \& Ryzhik(2007)]{jeffrey2007table}
{\sc \au{Gradshteyn, I.~S.} \& \au{Ryzhik, I.~M.}} \yr{2007} {\em Table of
  Integrals, Series, and Products\/}.  \publ{New York: Elsevier}.

\bibitem[Greenside \& Coughran~Jr(1984)]{greenside1984nonlinear}
{\sc \au{Greenside, H.~S.} \& \au{Coughran~Jr, W.~M.}} \yr{1984}
  \at{{Nonlinear pattern formation near the onset of Rayleigh-B{\'e}nard
  convection}}.  \jt{Phys. Rev. A}  \bvol{30}~(1),  \pg{398}.

\bibitem[Grossmann {\em et~al.\/}(2016)Grossmann, Lohse \& Sun]{Gro2016}
{\sc \au{Grossmann, S.}, \au{Lohse, D.} \& \au{Sun, C.}} \yr{2016}
  \at{High–{R}eynolds number {T}aylor-{C}ouette turbulence}.  \jt{Annu. Rev.
  Fluid Mech.}  \bvol{48}~(1),  \pg{53--80}.

\bibitem[Haan(1989)]{haan1989onset}
{\sc \au{Haan, S.~W.}} \yr{1989}  \at{{Onset of nonlinear saturation for
  Rayleigh-Taylor growth in the presence of a full spectrum of modes}}.
  \jt{Phys. Rev. A}  \bvol{39}~(11),  \pg{5812}.

\bibitem[Herminghaus {\em et~al.\/}(2014)Herminghaus, Maass, Kr{\"u}ger,
  Thutupalli, Goehring \& Bahr]{herminghaus2014interfacial}
{\sc \au{Herminghaus, S.}, \au{Maass, C.~C.}, \au{Kr{\"u}ger, C.},
  \au{Thutupalli, S.}, \au{Goehring, L.} \& \au{Bahr, C.}} \yr{2014}
  \at{Interfacial mechanisms in active emulsions}.  \jt{Soft matter}
  \bvol{10}~(36),  \pg{7008--7022}.

\bibitem[Hu {\em et~al.\/}(2019)Hu, Lin, Rafai \& Misbah]{PhysRevLettHu}
{\sc \au{Hu, W.}, \au{Lin, T.}, \au{Rafai, S.} \& \au{Misbah, C.}} \yr{2019}
  \at{Chaotic swimming of phoretic particles}.  \jt{Phys. Rev. Lett.}
  \bvol{123},  \pg{238004}.

\bibitem[Jeanneret {\em et~al.\/}(2016)Jeanneret, Pushkin, Kantsler \&
  Polin]{jeanneret2016entrainment}
{\sc \au{Jeanneret, R.}, \au{Pushkin, D.~O.}, \au{Kantsler, V.} \& \au{Polin,
  M.}} \yr{2016}  \at{Entrainment dominates the interaction of microalgae with
  micron-sized objects}.  \jt{Nat. Commun.}  \bvol{7}~(1),  \pg{1--7}.

\bibitem[Jiang {\em et~al.\/}(2010)Jiang, Chen, Tripathy, Luijten, Schweizer \&
  Granick]{jiang2010janus}
{\sc \au{Jiang, S.}, \au{Chen, Q.}, \au{Tripathy, M.}, \au{Luijten, E.},
  \au{Schweizer, K.~S.} \& \au{Granick, S.}} \yr{2010}  \at{Janus particle
  synthesis and assembly}.  \jt{Adv. Mater.}  \bvol{22}~(10),  \pg{1060--1071}.

\bibitem[Jin {\em et~al.\/}(2017)Jin, Kr{\"u}ger \& Maass]{jin2017chemotaxis}
{\sc \au{Jin, C.}, \au{Kr{\"u}ger, C.} \& \au{Maass, C.~C.}} \yr{2017}
  \at{Chemotaxis and autochemotaxis of self-propelling droplet swimmers}.
  \jt{Proc. N. Acad. Sci.}  \bvol{114}~(20),  \pg{5089--5094}.

\bibitem[Kr{\"u}ger {\em et~al.\/}(2016)Kr{\"u}ger, Kl{\"o}s, Bahr \&
  Maass]{kruger2016curling}
{\sc \au{Kr{\"u}ger, C.}, \au{Kl{\"o}s, G.}, \au{Bahr, C.} \& \au{Maass, C.}}
  \yr{2016}  \at{Curling liquid crystal microswimmers: A cascade of spontaneous
  symmetry breaking}.  \jt{Phys. Rev. Lett.}  \bvol{117}~(4),  \pg{048003}.

\bibitem[Lauga(2016)]{lauga2016bacterial}
{\sc \au{Lauga, E.}} \yr{2016}  \at{Bacterial hydrodynamics}.  \jt{Annu. Rev.
  Fluid Mech.}  \bvol{48},  \pg{105--130}.

\bibitem[Lauga \& Thomas(2009)]{lauga2009hydrodynamics}
{\sc \au{Lauga, E.} \& \au{Thomas, R.~P.}} \yr{2009}  \at{The hydrodynamics of
  swimming microorganisms}.  \jt{Rep. Prog. Phys.}  \bvol{72}~(9),
  \pg{096601}.

\bibitem[Li {\em et~al.\/}(2019)Li, Diddens, Prosperetti, Chong, Zhang \&
  Lohse]{li2019bouncing}
{\sc \au{Li, Y.}, \au{Diddens, C.}, \au{Prosperetti, A.}, \au{Chong, K.~L.},
  \au{Zhang, X.} \& \au{Lohse, D.}} \yr{2019}  \at{Bouncing oil droplet in a
  stratified liquid and its sudden death}.  \jt{Phys. Rev. Lett.}
  \bvol{122}~(15),  \pg{154502}.

\bibitem[Liu(2017)]{liu2017unified}
{\sc \au{Liu, L.~Q.}} \yr{2017} {\em Unified Theoretical Foundations of Lift
  and Drag in Viscous and Compressible External Flows\/}.  \publ{Singapore:
  Springer}.

\bibitem[Lohse \& Zhang(2020)]{lohse2020}
{\sc \au{Lohse, D.} \& \au{Zhang, X.}} \yr{2020}  \at{Physicochemical
  hydrodynamics of droplets out of equilibrium}.  \jt{Nat. Rev. Phys.}
  \bvol{2},  \pg{426--443}.

\bibitem[Long {\em et~al.\/}(1999)Long, Stone \&
  Ajdari]{long1999electroosmotic}
{\sc \au{Long, D.}, \au{Stone, H.~A.} \& \au{Ajdari, A.}} \yr{1999}
  \at{Electroosmotic flows created by surface defects in capillary
  electrophoresis}.  \jt{J. Colloid. Interf. Sci.}  \bvol{212}~(2),
  \pg{338--349}.

\bibitem[Maass {\em et~al.\/}(2016)Maass, Kr{\"u}ger, Herminghaus \&
  Bahr]{maass2016swimming}
{\sc \au{Maass, C.~C.}, \au{Kr{\"u}ger, C.}, \au{Herminghaus, S.} \& \au{Bahr,
  C.}} \yr{2016}  \at{Swimming droplets}.  \jt{Annu. Rev. Condens. Matter
  Phys.}  \bvol{7},  \pg{171--193}.

\bibitem[Michelin {\em et~al.\/}(2020)Michelin, Game, Lauga, Keaveny \&
  Papageorgiou]{michelin2020spontaneous}
{\sc \au{Michelin, S.}, \au{Game, S.}, \au{Lauga, E.}, \au{Keaveny, E.} \&
  \au{Papageorgiou, D.}} \yr{2020}  \at{Spontaneous onset of convection in a
  uniform phoretic channel}.  \jt{Soft Matter}  \bvol{16},  \pg{1259--1269}.

\bibitem[Michelin \& Lauga(2014)]{michelin2014phoretic}
{\sc \au{Michelin, S.} \& \au{Lauga, E.}} \yr{2014}  \at{Phoretic
  self-propulsion at finite {P}{\'e}clet numbers}.  \jt{J. Fluid Mech.}
  \bvol{747},  \pg{572--604}.

\bibitem[Michelin {\em et~al.\/}(2013)Michelin, Lauga \&
  Bartolo]{michelin2013spontaneous}
{\sc \au{Michelin, S.}, \au{Lauga, E.} \& \au{Bartolo, D.}} \yr{2013}
  \at{Spontaneous autophoretic motion of isotropic particles}.  \jt{Phys.
  Fluids}  \bvol{25}~(6),  \pg{061701}.

\bibitem[Moran \& Posner(2011)]{moran2011electrokinetic}
{\sc \au{Moran, J.~L.} \& \au{Posner, J.~D.}} \yr{2011}  \at{Electrokinetic
  locomotion due to reaction-induced charge auto-electrophoresis}.  \jt{J.
  Fluid Mech.}  \bvol{680},  \pg{31--66}.

\bibitem[Moran \& Posner(2017)]{moran2017phoretic}
{\sc \au{Moran, J.~L.} \& \au{Posner, J.~D.}} \yr{2017}  \at{Phoretic
  self-propulsion}.  \jt{Annu. Rev. Fluid Mech.}  \bvol{49},  \pg{511--540}.

\bibitem[Morozov \& Michelin(2019{\natexlab{{\em a\/}}})]{morozov2019nonlinear}
{\sc \au{Morozov, M.} \& \au{Michelin, S.}} \yr{2019{\natexlab{{\em a\/}}}}
  \at{Nonlinear dynamics of a chemically-active drop: From steady to chaotic
  self-propulsion}.  \jt{J. Chem. Phys.}  \bvol{150}~(4),  \pg{044110}.

\bibitem[Morozov \& Michelin(2019{\natexlab{{\em
  b\/}}})]{morozov2019orientational}
{\sc \au{Morozov, M.} \& \au{Michelin, S.}} \yr{2019{\natexlab{{\em b\/}}}}
  \at{Orientational instability and spontaneous rotation of active nematic
  droplets}.  \jt{Soft Matter}  \bvol{15}~(39),  \pg{7814--7822}.

\bibitem[Pearson(1958)]{pearson1958convection}
{\sc \au{Pearson, J. R.~A.}} \yr{1958}  \at{On convection cells induced by
  surface tension}.  \jt{J. Fluid Mech.}  \bvol{4}~(5),  \pg{489--500}.

\bibitem[Piazza(2008)]{piazza2008thermophoresis}
{\sc \au{Piazza, R.}} \yr{2008}  \at{Thermophoresis: moving particles with
  thermal gradients}.  \jt{Soft Matter}  \bvol{4}~(9),  \pg{1740--1744}.

\bibitem[van~der Poel {\em et~al.\/}(2015)van~der Poel, Ostilla-M{\'o}nico,
  Donners \& Verzicco]{van2015pencil}
{\sc \au{van~der Poel, E.~P.}, \au{Ostilla-M{\'o}nico, R.}, \au{Donners, J.} \&
  \au{Verzicco, R.}} \yr{2015}  \at{A pencil distributed finite difference code
  for strongly turbulent wall-bounded flows}.  \jt{Comput. Fluids}  \bvol{116},
   \pg{10--16}.

\bibitem[Qi {\em et~al.\/}(2020)Qi, Westphal, Gompper \&
  Winkler]{kai2020squimer}
{\sc \au{Qi, K.}, \au{Westphal, E.}, \au{Gompper, G.} \& \au{Winkler, R.~G.}}
  \yr{2020}  \at{Enhanced rotational motion of spherical squirmer in polymer
  solutions}.  \jt{Phys. Rev. Lett.}  \bvol{124},  \pg{068001}.

\bibitem[Ramaswamy(2010)]{ramaswamy2010mechanics}
{\sc \au{Ramaswamy, S.}} \yr{2010}  \at{The mechanics and statistics of active
  matter}.  \jt{Annu. Rev. Condens. Matter Phys.}  \bvol{1}~(1),
  \pg{323--345}.

\bibitem[Ruckenstein(1981)]{ruckenstein1981can}
{\sc \au{Ruckenstein, E.}} \yr{1981}  \at{Can phoretic motions be treated as
  interfacial tension gradient driven phenomena?}  \jt{J. Colloid. Interf.
  Sci.}  \bvol{83}~(1),  \pg{77--81}.

\bibitem[Spandan {\em et~al.\/}(2017)Spandan, Meschini, Ostilla-M{\'o}nico,
  Lohse, Querzoli, de~Tullio \& Verzicco]{spandan2017parallel}
{\sc \au{Spandan, V.}, \au{Meschini, V.}, \au{Ostilla-M{\'o}nico, R.},
  \au{Lohse, D.}, \au{Querzoli, G.}, \au{de~Tullio, M.~D.} \& \au{Verzicco,
  R.}} \yr{2017}  \at{A parallel interaction potential approach coupled with
  the immersed boundary method for fully resolved simulations of deformable
  interfaces and membranes}.  \jt{J. Comput. Phys.}  \bvol{348},
  \pg{567--590}.

\bibitem[Squires \& Bazant(2006)]{squires2006breaking}
{\sc \au{Squires, T.~M.} \& \au{Bazant, M.~Z.}} \yr{2006}  \at{Breaking
  symmetries in induced-charge electro-osmosis and electrophoresis}.  \jt{J.
  Fluid Mech.}  \bvol{560},  \pg{65--101}.

\bibitem[Suga {\em et~al.\/}(2018)Suga, S., Ichikawa \& Kimura]{suga2018self}
{\sc \au{Suga, M.}, \au{S., Suda}, \au{Ichikawa, M.} \& \au{Kimura, Y.}}
  \yr{2018}  \at{Self-propelled motion switching in nematic liquid crystal
  droplets in aqueous surfactant solutions}.  \jt{Phys. Rev. E}  \bvol{97}~(6),
   \pg{062703}.

\bibitem[Vajdi~Hokmabad {\em et~al.\/}(2019)Vajdi~Hokmabad, Baldwin, Kr\"uger,
  Bahr \& Maass]{babak2019}
{\sc \au{Vajdi~Hokmabad, B.}, \au{Baldwin, K.~A.}, \au{Kr\"uger, C.}, \au{Bahr,
  C.} \& \au{Maass, C.~C.}} \yr{2019}  \at{Topological stabilization and
  dynamics of self-propelling nematic shells}.  \jt{Phys. Rev. Lett.}
  \bvol{123},  \pg{178003}.

\bibitem[Vajdi~Hokmabad {\em et~al.\/}(2021)Vajdi~Hokmabad, Dey, Jalaal,
  Mohanty, Almukambetova, Baldwin, Lohse \& Maass]{babak2020}
{\sc \au{Vajdi~Hokmabad, B.}, \au{Dey, R.}, \au{Jalaal, M.}, \au{Mohanty, D.},
  \au{Almukambetova, M.}, \au{Baldwin, K.~A.}, \au{Lohse, D.} \& \au{Maass,
  C.~C.}} \yr{2021}  \at{Emergence of bimodal motility in active droplets}.
  \jt{Phys. Rev. X}  \bvol{11}~(1),  \pg{011043}.

\bibitem[Verzicco \& Orlandi(1996)]{verzicco1996finite}
{\sc \au{Verzicco, R.} \& \au{Orlandi, P.}} \yr{1996}  \at{A finite-difference
  scheme for three-dimensional incompressible flows in cylindrical
  coordinates}.  \jt{J. Comput. Phys.}  \bvol{123}~(2),  \pg{402--414}.

\bibitem[Wu {\em et~al.\/}(2006)Wu, Ma \& Zhou]{Wu2006}
{\sc \au{Wu, J.~Z.}, \au{Ma, H.~Y.} \& \au{Zhou, M.~D.}} \yr{2006} {\em
  Vorticity and Vortex Dynamics\/}.  \publ{Berlin: Springer}.

\end{thebibliography}
